\documentclass[12pt]{article}
\usepackage{amsmath}
\usepackage{mathtools} 
\usepackage[T1]{fontenc}
\usepackage{amssymb}
\usepackage{amsfonts}
\usepackage{amsthm}
\usepackage{mathrsfs}
\usepackage[all]{xy}
\usepackage{tensor}
\usepackage{comment}
\usepackage{stmaryrd}
\usepackage{bm}
\usepackage[normalem]{ulem}
\usepackage[usenames,dvipsnames]{xcolor}
\usepackage[colorlinks=true, citecolor=Blue, linkcolor=Blue]{hyperref}
\usepackage{ragged2e}
\usepackage{tikz}
    \usetikzlibrary{decorations.markings}

\usepackage{setspace}
\usepackage[labelfont={small, bf, sf}, textfont={small,sf}]{caption}
\numberwithin{equation}{section} 
\usepackage{authblk}
\usepackage[in]{fullpage}
\usepackage{cite}

\newcommand{\lie}{\pounds}
\newcommand{\ep}{\epsilon}

\newcommand{\h}[1]{{\hat{#1}}}

\newcommand{\beom}{\mathcal{E}} 

\newcommand{\sr}{\mathcal{U}}
\newcommand{\ns}{\mathcal{N}}
\newcommand{\ts}{\mathcal{T}}

\newcommand{\byc}{\mathcal{Q}^\text{BY}}
\newcommand{\ct}{\text{ct}}
\newcommand{\nn}[1]{\tilde{#1}}

\DeclareMathOperator{\dv}{div}

\newcommand{\ind}[1]{\indices{#1}}

\newcommand{\Hajicek}{H\'aj\'i\v{c}ek }

\newcommand{\beq}{\begin{equation}}
\newcommand{\eeq}{\end{equation}}
\newcommand{\bes}{\begin{subequations}}
\newcommand{\ees}{\end{subequations}}
\newcommand{\bea}{\begin{eqnarray}}
\newcommand{\eea}{\end{eqnarray}}

\newcommand{\be}{\begin{equation}}
\newcommand{\ee}{\end{equation}}

\title{Brown-York charges at null boundaries}
\author[1,2]{Venkatesa Chandrasekaran\thanks{venchandrasekaran@ias.edu}}
\author[3]{\'Eanna \'E. Flanagan\thanks{eef3@cornell.edu}}
\author[3]{Ibrahim Shehzad\thanks{is354@cornell.edu}}
\author[4,5]{Antony J. Speranza\thanks{asperanz@gmail.com}}
	
\affil[1]{\small \it Berkeley Center for Theoretical Physics, Berkeley, CA, 94720, USA}
\affil[2]{\small \it Institute for Advanced Study, Princeton, NJ, 08540, USA}
\affil[3]{\small \it Department of Physics, Cornell University, Ithaca, NY, 14853, USA}
\affil[4]{\small \it Perimeter Institute for Theoretical Physics, 31 Caroline St. N, Waterloo, ON N2L 2Y5, Canada}
\affil[5]{\small \it Department of Physics, University of Illinois, Urbana-Champaign, Urbana IL 61801, USA\vspace{-1.5cm}}
	
\date{}

\begin{document}
\maketitle

\begin{abstract}
The Brown-York stress tensor provides a means for defining quasilocal gravitational charges 
in subregions bounded by a timelike hypersurface.   We consider the generalization
of this stress tensor to null hypersurfaces.   Such a stress tensor can be derived from the on-shell
subregion action of general relativity associated with a Dirichlet variational principle, which
fixes an induced Carroll structure on the null boundary.   The formula for the mixed-index tensor 
$T{}^i{}_j$ takes a remarkably simple form that is manifestly independent of the choice of 
auxiliary null vector at the null surface, and we compare this expression
to previous proposals for null Brown-York stress tensors.  
The  stress tensor we obtain satisfies a covariant conservation
equation with respect to any connection induced from a rigging vector at the hypersurface, as a result of the null constraint equations.
For transformations
that act covariantly on the boundary structures, the Brown-York charges coincide with canonical charges constructed from a version of the Wald-Zoupas procedure.  
For anomalous transformations, the charges differ by an intrinsic functional of the boundary geometry,
which we explicity verify for a set of symmetries associated with finite null
hypersurfaces.  Applications of the null Brown-York stress tensor to 
symmetries of asymptotically flat spacetimes and celestial holography are discussed.  
\end{abstract}

\newpage
\tableofcontents

\section{Introduction}

Diffeomorphism invariance  is a defining feature of gravitational theories such as 
general relativity, 
giving rise to charges that  comprise 
an important set of observables in these theories.  
Although diffeomorphisms supported only in the bulk are pure gauge and hence associated with
 vanishing charges, transformations that act on the boundary of a spacetime manifold or 
subregion yield nontrivial charges 
that provide notions of energy and 
angular momentum in the region, including
contributions from the gravitational field. 
These charges have found applications in a number of 
recent works, including symmetries of asymptotically flat
space \cite{Barnich2010, Barnich:2011mi,  Strominger:2013jfa,
He:2014laa,Stro-lectures, Campiglia:2014yka, Campiglia:2015yka, 
Compere:2018ylh, Flanagan:2019vbl, freidel2021weyl},
asymptotic observables in holography and AdS/CFT \cite{Hollands:2005ya,
Papadimitriou:2005ii, Harlow:2019yfa}, soft hair for black hole horizons and its relation to the information problem 
\cite{Carlip_1999, Hawking_2016, Haco:2018ske, Chen:2020nyh, Chandrasekaran:2020wwn,
Flanagan:2021ojq, Flanagan:2021svq, Pasterski:2020xvn}, and edge modes and entanglement for subregions
\cite{Donnelly2016a, Geiller:2017xad, Speranza2018a, Geiller:2017whh, Geiller:2019bti, Freidel:2020xyx, Donnelly:2020xgu}.

When dealing with subregions bounded by
a finite, timelike hypersurface, the construction of 
Brown and York \cite{Brown:1992br} gives a prescription for determining the gravitational charges
in terms of the variational principle for the subregion.  By taking the on-shell
subregion action
$S^\text{cl}[h_{ij}]$ to be a functional of the induced metric $h_{ij}$ on the boundary, one can define 
a boundary stress tensor in the usual way as the functional derivative $T^{ij}
= \frac{2}{\sqrt{-h}}
\frac{\delta S^\text{cl}}{\delta h_{ij} }$.
Given an infinitesimal boundary diffeomorphism generated by a vector $\xi^i$, an associated 
boundary current can be formed 
using the stress tensor according to the formula $j_\xi = T\ind{^i_j} \xi^j \eta_i$,
where $\eta_i$ is the volume form on the boundary.  Integrating this current over a cut of the 
boundary yields the gravitational charge, and by choosing the vector field $\xi^i$ appropriately, 
one obtains in this way the Brown-York quasilocal energy and angular momentum.

These charges can be shown to agree with canonical charges generating the associated 
symmetry transformations on the gravitational phase space \cite{Hollands:2005ya, Papadimitriou:2005ii, Harlow:2019yfa}.  
This agreement holds when imposing Dirichlet boundary conditions to define
the subregion phase space, or more generally for charges constructed using the 
Wald-Zoupas procedure \cite{Wald:1999wa} with a Dirichlet form of the flux 
\cite{Chandrasekaran:2020wwn}.  
Since the Wald-Zoupas construction deals with  open Hamiltonian systems associated with 
subregions, the charges obtained are colloquially referred to as ``nonintegrable,''
which, more precisely, means that the transformations they generate on phase space do not 
reproduce the action of their associated diffeomorphism.  Generally, the Wald-Zoupas procedure 
suffers from a number of ambiguities related to the nonintegrability of Hamilton's equation
for the diffeomorphism transformation, but these ambiguities can be resolved by demanding
a Dirichlet form for the nonintegrable contribution 
\cite{Chandrasekaran:2020wwn, Chandrasekaran2021}.  Although the Brown-York charges 
appear to sidestep these subtleties involving integrability, the fact that they agree
with Wald-Zoupas charges with the Dirichlet flux condition demonstrates that they are 
simply employing the same resolution to the integrability problem.\footnote{In the Brown-York
context, instead of finding that Hamilton's equations are not integrable, one instead
sees that the subregion action is not stationary for perturbations involving nonzero 
$\delta h_{ij}$.  Demanding that the subregion action be stationary except for terms 
involving $\delta h_{ij}$ is equivalent to the Dirichlet flux condition for the 
Wald-Zoupas charges \cite{Chandrasekaran2021}.} 
In addition to providing a means for constructing 
canonical charges, the boundary stress tensor $T^{ij}$ also features prominently in holographic
dualities such as AdS/CFT, where it is interpreted as the stress tensor operator of the dual 
conformal field theory \cite{Balasubramanian:1999re, Myers:1999psa, DeHaro2001}. 

In many cases of interest, including exteriors of black hole event horizons, entanglement
wedges in holographic setups, and asymptotically flat spacetimes,
one is interested in subregions bounded by null hypersurfaces, as opposed to 
timelike ones. A natural question arises as to whether the Brown-York procedure
can be 
generalized to accommodate null hypersurfaces
in order to obtain gravitational charges  in this context.  
The goal of the present paper is to answer this question in the affirmative, and to derive 
an explicit expression for the null analog of the Brown-York stress tensor
for general relativity.  The stress tensor
has the surprisingly simply expression
\beq \label{eqn:mainresult}
T\ind{^i_j} = -\frac{1}{8\pi G} \left(W\ind{^i_j} - W \delta\ind{^i_j}\right),
\eeq
where the {\it shape operator} $W\ind{^i_j}$, defined in \eqref{eqn:Wdefn},
is the null surface analog of the mixed-index 
extrinsic curvature $K\ind{^i_j}$ of a timelike hypersurface.  
In fact, the null stress tensor (\ref{eqn:mainresult}) depends on $W\ind{^i_j}$ in precisely
the same way as the standard Brown-York stress tensor depends on $K\ind{^i_j}$, making the analogy
quite sharp. 

The expression (\ref{eqn:mainresult}) is obtained by considering the variational principle for 
general relativity in a subregion bounded by a null hypersurface. 
This variational principle requires a notion of Dirichlet boundary conditions for the 
null surface in order to write the subregion action as a functional of boundary geometric
data.  Unlike the timelike case, where the intrinsic geometry is naturally that of 
a pseudo-Riemannian structure associated with the induced metric $h_{ij}$, there are 
a number of different choices for how to define the intrinsic quantities of the null surface 
that are fixed in a Dirichlet variational principle.  
The choice leading to (\ref{eqn:mainresult}) comes from imbuing the null boundary
with a Carrollian structure, consisting of a degenerate metric $q_{ij}$ and a preferred null
generator $n^i$ satisfying $n^i q_{ij}=0$
\cite{Henneaux1979, Duval:2014lpa, Duval:2014uoa, Duval:2014uva, Henneaux:2021yzg}. 
This structure arises naturally from 
the spacetime geometry after fixing a preferred  normal $n_a$ to the null surface.  
The variational principle for general relativity with this boundary condition was explored 
extensively in \cite{Chandrasekaran:2020wwn}, 
and utilizes a null analog of the Gibbons-Hawking-York 
boundary term that has appeared in a number of recent works 
\cite{Parattu_2016, Lehner_2016, Hopfmuller2017a, Oliveri:2019gvm,Aghapour:2018icu}.

Previously there have been two other proposals for a null Brown-York stress
tensor, put forward by Jafari \cite{Jafari:2019bpw} and Donnay and Marteau \cite{Donnay2019},  
each of which differs
 from the expression (\ref{eqn:mainresult}).  The  discrepancies are due to the 
different choices of geometric structures to associate with the null surface 
and the corresponding differences in boundary conditions to employ when defining the 
subregion variational principle.  Jafari's construction
utilizes a spacelike foliation of the null surface in lieu of a preferred normalization
of the null generator.  Donnay and Marteau obtain their stress tensor  using a null-limit 
of timelike hypersurfaces, which induces a scalar function on the null surface
that can be interpreted as a local surface gravity.  In section \ref{sec:compareby}, we  
describe the precise relation between the different null Brown-York stress tensors, and 
examine how they arise from these different choices of geometric structures on the null
surface.

The demonstration of the equivalence between Brown-York and canonical charges requires the 
stress tensor to satisfy a  conservation equation.  
In the timelike case, 
this conservation equation simply states that the stress tensor is divergenceless with respect
to the unique connection $D_i$ compatible with the induced metric.  Null surfaces are 
more subtle in this regard, since a Carrollian structure does not determine a unique connection
with respect to which to define the covariant conservation of $T\ind{^i_j}$.  Nevertheless,
we show in section \ref{sec:geonullsurfaces}
that there is a class of torsion-free, but generically not metric compatible,
connections associated with the Carroll structure
that are naturally induced from the spacetime Levi-Civita connection as
{\it rigged connections}, using a construction of Mars and Senovilla \cite{Mars1993}. 
While such connections have appeared previously in describing null infinity and 
finite null surfaces embedded in spacetime
\cite{Geroch-asymp,Ashtekar:1987tt,CFP}, 
they have not been considered
 in the recent literature on Carroll geometry (see, however, \cite{Dautcourt1967}), and we comment on their main 
properties in appendix \ref{app:carroll}.  
We further show that the conservation of the stress tensor with respect to any such 
connection is equivalent to the constraint equations of general relativity on the null surface,
consisting of the Raychaudhuri and Damour-Navier-Stokes equations 
\cite{Gourgoulhon:2005ng, Damour1982}.  The connection between the 
gravitational constraint equations and conservation laws was also explored by Donnay and Marteau
for their null-limit stress tensor
\cite{Donnay2019}, and in section \ref{sec:compareby} 
we compare their conservation equation
to the one obtained for the stress tensor in the present work.  

After demonstrating the equivalence between the null Brown-York charges and canonical charges
for transformations that act covariantly on
the Carrollian geometry, we turn our attention in
section \ref{sec:byanom} to so-called anomalous transformations.  These arise from bulk diffeomorphisms
that do not fully preserve the fixed null normal $n_a$, and were shown in \cite{Chandrasekaran:2020wwn} to be
the essential feature determining extensions of gravitational charge algebras, which,
when evaluated on a black hole horizon, lead to information about the horizon entropy.  
We demonstrate that for such anomalous transformations, the Brown-York and canonical charges 
in general
do no agree, instead differing by a functional of the intrinsic geometry.  
We explicitly exhibit this difference by comparing the Brown-York expression to 
the canonical charges associated with BMS-like transformations on finite null surfaces that 
were obtained by Chandrasekaran, Flanagan, and Prabhu 
\cite{CFP}.  

We conclude in section \ref{sec:discussion} 
with some discussions on potential applications
to symmetries of asymptotically flat space, celestial holography, and the fluid-gravity 
correspondence, and comment on some directions for future work.

\subsection{Notation}
 Latin letters from the beginning of the alphabet $a,b,c,\ldots$ are used to denote 
spacetime tensor indices, while those from the middle of the alphabet $i,j,k,\ldots$ 
are used for tensors defined on a timelike or null bounding hypersurface.  
Differential forms such as the spacetime volume form $\ep$ or hypersurface volume 
form $\eta$ are often written with indices suppressed.  When denoting a contraction 
on one or more indices, we will use the shorthand $\ep_a$ to indicate the indices which 
are contracted, while continuing to suppress the remaining indices.  We also use the notation
$i_V$ for contraction with a vector $V^a$ into a differential form.

\section{Timelike boundary}

We begin by reviewing the construction of Brown-York charges for timelike boundaries, and 
the argument demonstrating their equivalence to canonical charges. 
This argument is familiar from previous considerations
regarding so-called ``counterterm subtraction charges'' in asymptotically anti-de Sitter
spaces 
\cite{Hollands:2005ya, Papadimitriou:2005ii}
and also discussions of integrable charges 
for finite timelike boundaries \cite{Harlow:2019yfa}.

Given an open subregion $\sr$ bounded in spatial extent by a timelike surface $\ts$, the action
for the subregion is given by a sum of a bulk Einstein-Hilbert term and the boundary 
Gibbons-Hawking-York term,\footnote{We leave
out contributions from future or past boundaries and codimension-2 corners, 
which are not needed in obtaining the 
Brown-York charges.}
\begin{align}
S &= \int_{\sr} L - \int_{\ts} \ell \\
L &= \frac{1}{16\pi G}(R-2\Lambda) \epsilon \label{eqn:EH} \\
\ell &= -\frac{1}{8\pi G} K\eta, \label{eqn:GHY}
\end{align}
where $\epsilon$ is the spacetime volume form, $R$ is 
the spacetime Ricci scalar, and $K$ is the trace of the extrinsic curvature of $\ts$.  Also $\eta$ is the induced volume form on $\ts$, defined such that  
$\ep \overset{\ts}{=} -n \wedge \eta$,
where $n_a$ is the outward pointing unit normal
to $\ts$.
The orientation of $\ts$ is chosen to be that determined by $\eta$.\footnote{This 
sign convention for $\eta$ is opposite to that used in reference
\cite{Harlow:2019yfa}, and is opposite the orientation induced on $\ts$ when
viewed as a component of $\partial\sr$. This implies that Stokes theorem
for the subregion takes the form $\int_\sr d\alpha = -\int_\ts \alpha$ (dropping
contributions from other components of $\partial\sr$).  This orientation for 
$\ts$ ensures that Stokes theorem for integrals over a segment of $\ts_1^2$ between
two cuts $\mathcal S_1$ and $\mathcal S_2$ of $\ts$, with $\mathcal S_1$ to the past of 
$\mathcal S_2$ takes the form 
$\int_{\ts_1^2} d\alpha = \int_{\mathcal S_2} \alpha - \int_{\mathcal S_1} \alpha$,
where the orientations of $\mathcal S_{1,2}$ are those induced by viewing them
as the boundary of bulk spacelike hypersurfaces $\Sigma_{1,2}$.  
Hence, if we take $\sr_1^2$ to be the region of $\sr$ bounded spatially by 
$\ts_1^2$ and to the past and future by $\Sigma_{1,2}$, these conventions 
imply that $\partial \sr_1^2 = -\ts_1^2 + \Sigma_2 - \Sigma_1$ (where the sign
indicates the relative orientations), $\partial \Sigma_{1,2} = \mathcal{S}_{1,2}$,
and $\partial \ts_1^2 = \mathcal{S}_2 - \mathcal{S}_1$.}  
The boundary term $\ell$ is chosen so that the action satisfies a 
Dirichlet variational principle with respect to the induced
metric $h_{ij}$ on $\ts$.  Its variation is given by (see e.g. 
\cite{Harlow:2019yfa, Burnett1990})
\beq \label{eqn:delS}
\delta S =\int_{\sr} E^{ab}\delta g_{ab} + \int_\ts\Big(\pi^{ij}\delta h_{ij} - d\beta\Big),
\eeq
where $E^{ab}=0$ are the vacuum Einstein field equations, the momenta $\pi^{ij}$ are given by
\beq
\pi^{ij} = -\frac{1}{16\pi G}(K^{ij} - K h^{ij})\eta,
\eeq
and $\beta$ contributes terms that localize to the past and future boundaries of 
$\ts$; explicitly, it is given by 
\beq \label{eqn:beta}
\beta = \frac{1}{16\pi G}\left(g^{ab} - n^a n^b \right)n^c\delta g_{bc} \eta_a,
\eeq
although we will drop these terms since we are ignoring contributions from future and 
past boundaries.

The action $S$ is therefore stationary\footnote{Up to contributions from future and past boundaries.}
when the bulk equations of motion hold and the induced metric $h_{ij}$ is fixed.  This then
 allows us to view the classical action as a functional of $h_{ij}$, $S^\text{cl}[h_{ij}]$, 
and the boundary stress tensor is given by the variation of this functional with 
respect to $h_{ij}$, 
\beq \label{eqn:TBYdefn}
T^{ij} = \frac{2}{\sqrt{-h}} \frac{\delta S^\text{cl}}{\delta h_{ij}} 
=  -\frac{1}{8\pi G}\left(K^{ij} - K h^{ij} \right).
\eeq
This stress tensor can be used to construct boundary Noether currents associated with 
infinitesimal diffeomorphisms that preserve the hypersurface $\ts$.  These are 
generated by vector fields $\xi^a$ tangent to $\ts$, and if $\xi^i$ is the restriction
of the vector to $\ts$, the current is given 
by\footnote{The sign in this equation is somewhat nonstandard, and arises due to the 
choice of orientation of $\ts$.  The stress tensor (\ref{eqn:TBYdefn})
is insensitive to the choice of orientation, since it arises from the 
$\int_\ts \pi^{ij}\delta h_{ij}$ term in (\ref{eqn:delS}) after stripping off the volume 
form $\eta$ and the integral over $\ts$.  Since under a change in orientation
$\eta\rightarrow -\eta$ and $\int_\ts \rightarrow -\int_\ts$, these signs 
cancel out in the definition of $T^{ij}$. However, the current $j_\xi$ is constructed
by contracting with the volume form $\eta$, and hence it flips sign under a change in
orientation.  As we will see below, the current that reproduces the canonical 
charges is the one associated with the orientation on $\ts$ naturally induced 
as a component of $\partial\sr$.  The volume form for this induced 
orientation is $-\eta$, which accounts for the sign in (\ref{eqn:jxiT}).
\label{ftn:orientation}}
\beq\label{eqn:jxiT}
j_\xi = -T\ind{^i_j} \xi^j \eta_i.
\eeq

The Noether current is conserved when $\xi^i$ generates a symmetry of the boundary
metric, which can be seen by computing its divergence,
\beq\label{eqn:djxi}
dj_\xi = -\eta( T^{ij}D_i \xi_j + D_iT\ind{^i_j} \xi^j) = 
-\frac\eta2 T^{ij}\lie_\xi h_{ij} -
(\dv T)_j \xi^j \eta,
\eeq
where $D_i$ is the connection compatible with $h_{ij}$.
The second term in this expression involving $\dv T$ is recognized as the momentum constraint
of general relativity associated with the hypersurface $\ts$, and hence vanishes on-shell.  
This is a feature that will continue to hold in the null case: the covariant conservation
of the Brown-York stress tensor is equivalent to imposing the momentum constraints of vacuum
general relativity on $\ts$.  When matter is present, there would be an additional boundary
contribution from the matter stress tensor so that the total boundary stress tensor
is conserved.  The 
first term in (\ref{eqn:djxi}) clearly vanishes when $\xi^i$ is a Killing vector for $h_{ij}$, 
and hence in this case 
$j_\xi$ defines a conserved current.  

Even when $\xi^i$ does not generate a symmetry of the boundary metric, the current $j_\xi$ defines
an important quantity due to its relation to gravitational charges constructed using 
canonical methods.  In particular, the Wald-Zoupas procedure 
\cite{Wald:1999wa, CFP,GPS}
employs covariant phase space techniques to construct charges that generically
are not conserved due to the presence of fluxes through the subregion boundary $\ts$.  
Instead of a conservation equation, these charges satisfy a continuity equation where 
the change in the charge is related to a well-defined flux.  Utilizing the reformulation 
of the Wald-Zoupas procedure given in references 
\cite{Chandrasekaran:2020wwn, Chandrasekaran2021},
which applies the techniques developed by Harlow and Wu 
\cite{Harlow:2019yfa}  for dealing with 
boundaries in the covariant phase space, the gravitational charges can be defined as the 
integral of a current $h_\xi$ over a cut of the boundary $\ts$.  The explicit expression
for $h_\xi$ is 
\beq \label{eqn:hxi}
h_\xi = Q_\xi +i_\xi \ell - \beta[\lie_\xi g_{ab}],
\eeq
where 
\beq
Q_\xi = -\frac{1}{16\pi G} \ep_{ab}\nabla^a \xi^b
\eeq
is the Noether potential, $\ell$ is the boundary term (\ref{eqn:GHY}), and $\beta[\lie_\xi g_{ab}]$
is the quantity (\ref{eqn:beta}) evaluated with $\delta g_{ab} = \lie_\xi g_{ab}$.  
Using that the boundary term $\ell$ transforms covariantly under any diffeomorphism generated by 
a vector field $\xi^a$ that is tangent to $\ts$, one can derive a continuity equation for $h_\xi$
of the form
\beq
d h_\xi  = -\pi^{ij} \lie_\xi h_{ij},
\eeq
where the expression on the right represents the flux density through the boundary $\ts$.

Since $\pi^{ij} = \frac12 \eta T^{ij}$, comparing to equation (\ref{eqn:djxi}) and imposing the 
constraint $\dv T = 0$ allows us to conclude 
\beq
dh_\xi = dj_\xi.
\eeq
Hence the charge densities $h_\xi$ and $j_\xi$ can differ at most by 
a closed form $s_\xi$.  
Furthermore, since $h_\xi$ and $j_\xi$ are covariantly
constructed from $\xi^i$ and the boundary fields for arbitrary choices of $\xi^i$, 
we can apply a theorem due to Wald \cite{W-closed} to conclude that 
$s_\xi$ is exact, $s_\xi = d c_\xi$.  This then implies that the charges
obtained by integrating the charge densities over a cut ${\cal S}$ of $\ts$ are 
insensitive to the choice of $c_\xi$, implying that the 
Wald-Zoupas and Brown-York charges coincide,
\beq \label{eqn:HxiQxiBY}
H_\xi = \int_{{\cal S}} h_\xi = \int_{{\cal S}}j_\xi \equiv \byc_\xi.
\eeq

Finally, 
as discussed in Refs. 
\cite{Witten:1998qj, Balasubramanian:1999re, DeHaro2001, Papadimitriou:2005ii,
MM-term,
Chandrasekaran2021},
when taking a limit of the surface $\ts$ to an asymptotic boundary, 
additional boundary terms 
$\ell_\ct$ must be added 
to the action in order to obtain finite charges 
in a process known as holographic renormalization.  These 
additional counterterms are required to be covariant functionals of the intrinsic
geometry in order to preserve the Dirichlet form of the variational principle. 
Such terms then change $\pi^{ij}$ by the variational derivative 
$\pi^{ij}_\ct = \frac{\delta\ell_\ct}{\delta h_{ij}}$.  
Covariance of $\ell_\ct$ is crucial for ensuring that the 
new stress tensor continues to be conserved, which requires 
$D_i \pi^{ij}_\ct = 0$.  This conservation equation is simply the Noether 
identity associated with the covariant functional $\ell_\ct$, which holds 
identically due to invariance of $\ell_\ct$ under boundary diffeomorphisms
\cite{KS2011, Jacobson:2011cc, Seifert:2006kv}.
Hence, we can conclude that the equality of Wald-Zoupas and Brown-York charges
is unaffected by the process of holographic renormalization, provided the 
the boundary counterterms are covariant functionals of the intrinsic boundary 
geometry.\footnote{These anomalous transformations manifest as a mismatch between 
the transformation of $\ell_\text{ct}$ on the phase space and its transformation
under the Lie derivative, $\delta_\xi \ell_\ct \neq \lie_\xi \ell_\text{ct}$.}
Note that subtleties can arise for transformations that are only tangential to $\ts$ asymptotically,
as occurs for some asymptotic symmetry transformations.  These can induce anomalous transformations
on $\ell_\text{ct}$ and the intrinsic quantities on $\ts$, due to the nonzero transverse
component of the vector field.   In these cases, the canonical 
and Brown-York charges can differ by terms related to holographic Weyl anomalies
\cite{Witten:1998qj,Henningson1998a, Balasubramanian:1999re, DeHaro2001,
Papadimitriou:2005ii, Hollands:2005wt}.  
A related example of this effect that occurs for null surfaces is examined in section 
\ref{sec:byanom}.

\section{Null boundary} \label{sec:nullbdy}

We can now repeat this analysis for a subregion bounded by
a null hypersurface $\ns$.  The main subtlety is that 
the intrinsic geometry is no longer characterized by a pseudo-Riemannian structure,
involving a nondegenerate metric.  
Instead, the geometry consists of a degenerate metric $q_{ij}$ with a single null
direction, and a preferred generator $n^i$ of the kernel of $q_{ij}$, i.e.\ a vector satisfying 
$n^i q_{ij}=0$.  Together, these objects define
a (weak) Carroll structure for the null surface $\ns$
\cite{Henneaux1979, Duval:2014lpa, Duval:2014uoa, Duval:2014uva, Henneaux:2021yzg}.  
While the degenerate metric $q_{ij}$ is naturally induced as the pullback
of the spacetime metric to $\ns$, the additional structure encoded in $n^i$
arises  after choosing 
a distinguished normal form $n_a$ of the null surface, after which $n^i$ is determined 
by raising the index with $g^{ab}$ and restricting the vector to $\ns$.  These structures were
argued in \cite{Chandrasekaran:2020wwn} to be the natural quantities with respect to which to formulate
the Dirichlet variational principle for null boundaries.  

There are two main subtleties associated with working with 
a Carrollian, as opposed to a pseudo-Riemannian, structure.  
First, the process of  lowering
indices with the degenerate intrinsic metric 
does not produce an isomorphism between tangent and cotangent 
vectors, and hence the index placement for tensors becomes important.  
Second,  there is no preferred connection
available for defining covariant derivatives of tensors.  
  Despite these complications,
we will find that a null version 
of the Brown-York stress tensor $T\ind{^i_j}$ can 
be obtained independent of any choice of connection.  It 
is naturally defined with one contravariant and one covariant index,
as is appopriate when viewing the stress tensor 
as a linear map from vectors $\xi^i$ into their
associated charge densities 
$j_\xi = -T\ind{^i_j}\xi^j \eta_i$, with
$\eta_i$ the volume form
\cite{deBoer:2017ing}.  Furthermore, a connection-independent notion
of the covariant conservation of the stress tensor $(\dv T)_j = 0$ also arises from the 
definition, and this condition turns out to precisely coincide with the 
imposition of the constraint equations on the null surface.

\subsection{Geometry of null surfaces}
\label{sec:geonullsurfaces}

In order to describe the Dirichlet variational principle with a null boundary, we need 
to review a few details on the intrinsic and extrinsic geometry of null surfaces.
The degenerate metric $q_{ij}$ determines a spatial volume form $\mu$ (up to a sign, which
can be fixed by a choice of orientation), which is 
a horizontal form of maximal degree, meaning it is one degree below a top form
and $i_n\mu = 0$.  This spatial volume form is such that on any codimension-1 cut
of $\ns$, $\mu$ pulls back to the induced volume form compatible with the 
pullback of $q_{ij}$ and the chosen orientation of the cut. The preferred null generator $n^i$ also determines a volume
form $\eta$ on the full null surface, which is the unique top form that 
satisfies $i_n \eta = \mu$.  These structures then determine a set of 
first order differential invariants, the expansion $\Theta$, extrinsic curvature 
$K_{ij}$,\footnote{Despite this terminology, the ``extrinsic curvature'' $K_{ij}$ 
is fully determined by the intrinsic quantities $(q_{ij}, n^i)$.} and shear $\sigma_{ij}$ according to the equations
\begin{align}
d\mu &= \Theta\eta \\
K_{ij} &= \frac12 \lie_n q_{ij} \\
\sigma_{ij} &= K_{ij} - \frac{1}{n-1} \Theta q_{ij}.
\end{align}
where $n = d-1$ is the dimension of the null hypersurface.

The extrinsic geometry is characterized by the {\it shape operator}
or {\it Weingarten map} $W\ind{^i_j}$ of the null surface,
which is determined after fixing a preferred spacetime 1-form $n_a$ at $\ns$ to serve as the null 
normal. Letting $\Pi\ind{^a_i}$ denote the pullback map to $\ns$, we note that 
the spacetime covariant derivative of the null normal $\nabla_a n^b$ upon taking a pullback 
produces a tensor 
$\Pi\ind{^a_i} \nabla_a n^b$ whose $b$ index is tangential.  This therefore defines 
a tensor $W\ind{^i_j}$ on $\ns$ which we refer to as the shape operator; explicitly
it can be defined as the unique tensor satisfying
\beq\label{eqn:Wdefn}
W\ind{^i_j}\Pi\ind{^b_i} = \Pi\ind{^a_j}\nabla_a n^b.
\eeq
Some components of the shape operator are determined by the intrinsic geometry
of $\ns$; in particular, the extrinsic curvature is obtained by lowering an index 
with $q_{ij}$,
\beq\label{eqn:WK}
W\ind{^i_j}q\ind{_i_k} = K_{kj}.
\eeq
The remaining components involve terms in $W\ind{^i_j}$ of the form $n^i \rho_j$, which
do not contribute to equation (\ref{eqn:WK}).  One such component arises from the 
equation
\beq\label{eqn:inaffinity}
W\ind{^i_j} n^j = k n^i,
\eeq
which holds since $n^a$ is parallel to null geodesics in spacetime.  Hence,
$n^i$ is an eigenvector of the shape operator, and its eigenvalue $k$ is called the 
inaffinity.  

To be more explicit about the decomposition of the shape operator, we need to 
introduce an auxiliary one-form $l_i$, normalized relative to the null generator
by $n^il_i = -1$.  Note that there is no preferred choice for $l_i$ for a generic
null surface.  Intrinsically, $l_i$ defines an Ehreshmann connection  \cite{Bekaert:2015xua, Ciambelli2019b}, which is just a projector $s\ind{^i_j} = -n^i l_j$ onto vertical
vectors parallel to $n^i$.  This similarly allows us to define a projector $q\ind{^i_j}$
onto horizontal forms through the relation
\beq
q\ind{^i_j} = \delta\ind{^i_j} + n^i l_j,
\eeq
as well as a partial inverse $q^{ij}$ of the degenerate metric $q_{jk}$ through the 
relations $q^{ij}q_{jk} = q\ind{^i_k}$, $q^{ij} l_j = 0$.
From the extrinsic perspective, $l_i$ can be taken to arise from 
a null rigging vector, which is a transverse vector $l^a$ defined on the null
surface satisfying $l\cdot n = -1$, $l \cdot l = 0$.  
This implies that $l^a$ is an outward pointing vector at the null
surface.  
Then $l_i = \Pi^{a}{}_i l_a$ defines
the desired one-form on the null surface.  
Having introduced $l_i$, the shape operator can then be decomposed as
\begin{align}
W\ind{^i_j} &= K\ind{^i_j} + n^i\rho_j \label{eqn:Wdecomp}\\
\rho_j &= \varpi_j -kl_j
\end{align}
where $K\ind{^i_j} = q^{ik}K_{kj}$, 
$\rho_j$ is the rotation one-form defined by, 
\beq
\rho_j = -\Pi\ind{^a_j} l_b\nabla_a n^b,
\eeq
and the \Hajicek one-form $\varpi_i$ is the spatial projection 
of $\rho_j$,
\beq
\varpi_i = q\ind{^j_i} \rho_j.
\eeq
We can also explicitly express the volume form on $\ns$ in terms of $l^a$ and the 
spacetime volume form by writing $\eta = \Pi^*(-i_l \ep)$, where $\Pi^*$ indicates
a pullback.  Note as in the timelike case, the volume form $\eta$ is associated 
with an orientation of $\ns$ that is opposite the natural orientation 
induced on $\ns$ as a component of the boundary of $\sr$.

The rigging vector $l^a$ also provides a natural projector for spacetime vectors
onto the null surface, given by
\beq
\Pi\ind{^a_b} = \delta\ind{^a_b} + l^a n_b,
\eeq
which then allows us to define an inclusion map $\Pi\ind{^i_a}$ that inverts the pullback
map for covectors in the sense
\beq
\Pi\ind{^i_b} \Pi\ind{^a_i} = \Pi\ind{^a_b}.
\eeq
We can then use $\Pi\ind{^i_a}$ and $\Pi\ind{^b_j}$ 
to map intrinsic tensor fields on the 
null surface into spacetime tensor fields defined at $\ns$.

Additionally, this projector  induces a natural rigged connection on the null surface 
from the spacetime Levi-Civita connection
through a construction of Mars and Senovilla 
\cite{Mars1993}.  
If $V^a$ is tangent to $\ns$, we can define the rigged covariant derivative 
as 
\beq \label{eqn:DaVb}
D_a V^b = \Pi\ind{^c_a}\Pi\ind{^b_d}\nabla_c V^d,
\eeq
which then defines an intrinsic covariant derivative on the vector 
$V^i = \Pi\ind{^i_a}V^a$ to be
\beq \label{eqn:DiVj}
D_i V^j = \Pi\ind{^a_i}\Pi\ind{^j_b}D_a V^b.
\eeq
$D_i$ is extended to covectors $U_i$ by first mapping it to spacetime using the inclusion map
$\Pi\ind{^i_a}$, taking the covariant derivative, and then pulling back,
\beq
D_i U_j = \Pi\ind{^a_i}\Pi{^b_j} \nabla_a(\Pi\ind{^k_b} U_k).
\eeq
$D_i$ is then extended in the usual way to tensors of arbitrary degree.  
It is important to emphasize that this intrinsic connection $D_i$ 
depends on the choice of auxiliary one-form $l_i$.
Note that it is manifestly torsion-free, but generically
does not preserve any of the intrinsic structures on the surface.  Instead, we 
have the following relations (derived in appendix \ref{app:carroll}): 
\begin{align}
D_i q_{jk} &= l_j K_{ik} + l_k K_{ij} \label{eqn:Dq}\\
D_i n^j &= W\ind{^j_i} = K\ind{^j_i} + n^j \rho_i \label{eqn:Dn}\\
D_i\eta &=-\rho_i \eta \label{eqn:Deta}\\
D_i \mu &= K\ind{^j_i}\eta_j  \label{eqn:Dmu}
\end{align}
From the intrinsic perspective, these relations can also be used as the 
definition of a connection compatible with a given Carroll structure 
and associated Ehresmann connection $l_i$.  In doing so, the rotation one-form
$\rho_i$ appears as additional data needed to fully specify the connection,
beyond that contained in $(q_{ij}, n^i, l_j)$.  Additionally, there is a final
relation involving $D_i l_j$ that is not fixed by equations (\ref{eqn:Dq}-\ref{eqn:Dmu}),
as shown in equation (\ref{eqn:Dl}).  The data in $D_i l_j$ that is not fixed by 
quantities already defined is captured by its symmetric, horizontal component,
$\nu_{ij} = q\ind{^m_i} q\ind{^n_j} D_{(m} l_{n)}$.  
As equation (\ref{eqn:Deta}) 
shows, $\rho_i$ characterizes the failure of the connection to preserve the volume form.
Similarly, $\nu_{ij}$ measures the failure of the connection to preserve the Ehresmann connection
$l_i$, although there are additional obstructions to the vanishing of $D_i l_j$
described in appendix \ref{app:carroll}.  
Intrinsically, one is free to work with a connection that imposes $\rho_i=0$
and $\nu_{ij}=0$; however, 
there are preferred, generically nonzero, choices for these quantities when working 
with a rigged connection induced from the spacetime connection.  

Finally, we mention how to express the divergence of a vector field 
on the null surface in terms of the connection $D_i$.  Since the null surface 
has a preferred volume form $\eta$, the divergence of a vector field $V^i$can be 
defined independently of a connection through the equation
\beq
d i_V\eta = (\dv V) \eta.
\eeq
Because the connection $D_i$ generally does not preserve the volume 
form, the expression for $\dv V$ in terms of $D_i$ contains a contribution
from $\rho_i$,
\beq\label{eqn:nulldiv}
\dv V = D_i V^i - \rho_i V^i.
\eeq

\subsection{Null Brown-York stress tensor} \label{sec:nullby}

We can now describe the construction of the null boundary Brown-York stress
tensor, and demonstrate the equivalence between the charges constructed from
it and the canonical charges.  
Before deriving the result, we first comment on an important point regarding the 
index placement of the stress tensor we are seeking to obtain.  On timelike surfaces,
the presence of a nondegenerate metric allows indices to be raised and lowered, and 
so the tensors $T_{ij}$, $T^{ij}$ and $T\ind{^i_j}$ all contain the same information. 
This is no longer true on a null surface, and there is a question as to which index placement
is correct.  
The answer is that the stress tensor is naturally defined as a mixed index object, $T\ind{^i_j}$.
This is because the stress tensor should be viewed as a map from a vector field $\xi^i$ to an
associated current $j_\xi$, which can be integrated over codimension-1 surfaces inside of 
$\ns$ to obtain fluxes of energy and momentum.  This current is obtained by contracting the 
vector $T\ind{^i_j}\xi^j$ into the volume form $\eta$ on $\ns$ (see 
related comments in \cite{deBoer:2017ing}).  
Note that the presence of a volume form $\eta$ as a natural structure characterizing 
the geometry of $\ns$ is important for obtaining a two-index tensor.  
This suggests that an even more natural object characterizing the stress-energy of the 
theory is the covector-valued differential form $T\ind{^i_j} \eta_i$.
However, since there is a preferred volume form when working with a fixed null normal 
$n_a$, we will focus on the associated stress tensor $T\ind{^i_j}$.

We begin as before with an open subregion $\sr$ in spacetime, now bounded in spatial extent
by a null hypersurface $\ns$. Nullness of $\ns$ is imposed as a boundary condition for the 
phase space of field configurations, and we further impose that $\ns$ be equipped with 
a preferred null normal that is also fixed when taking variations, $\delta n_a=0$.
The action is taken to contain 
the same bulk term (\ref{eqn:EH}), but in place of the 
Gibbons-Hawking-York term, the boundary term for the null surface is constructed 
from the inaffinity $k$ according to\footnote{Another choice for the boundary term is $-\frac{1}{8\pi G}(k+\Theta)\eta$ 
\cite{Parattu_2016, Aghapour:2018icu, Oliveri:2019gvm}.
The additional term involving $\Theta$ 
can be shown to be a total derivative on $\ns$, and hence only changes the 
definition of 
the corner term $\beta$ (see \eqref{eqn:nullcorner}), and does not affect the Dirichlet variational principle.}
\beq \label{eqn:nullbdy}
\ell = -\frac{1}{8\pi G} k \eta,
\eeq
and the subregion action is defined to be 
\beq
S = \int_\sr L - \int_\ns \ell,
\eeq
where the orientation of $\ns$ is again chosen to be opposite the induced 
orientation as a component of $\partial \sr$. 
The variation of the action with this boundary term takes the form of a Dirichlet 
variational principle \cite{Chandrasekaran:2020wwn}, 
\beq \label{eqn:delSnull}
\delta S = \int_{\sr} E^{ab}\delta g_{ab} 
+ \int_\ns\Big( \pi^{ij}\delta q_{ij} + \pi_i \delta n^i
- d\beta\Big)
\eeq
where 
\begin{align}
\pi^{ij} &= -\frac{1}{16\pi G}\left(K^{ij} - (\Theta + k) q^{ij}\right) \eta \label{eqn:piij}\\
\pi_i &= \frac{1}{8\pi G} \left(\Theta l_i+\varpi_i \right) \eta \label{eqn:pii}\\
\beta &= \frac{1}{16\pi G}\left(g^{ab} n^c - n^a g^{bc}\right)\delta g_{ab} \eta_c.
\label{eqn:nullcorner}
\end{align}
This action is stationary when the bulk equations of motion hold and the intrinsic Carroll
structure defined by $(q_{ij}, n^i)$ is held fixed, which allows the classical action to 
be viewed as a functional of this structure, $S^\text{cl}[q_{ij}, n^i]$.

Since the classical action is now a functional of two geometric quantities $(q_{ij}, n^i)$ 
instead of a single metric $h_{ij}$, defining a stress tensor associated with 
it requires slightly more care than in the timelike case.    
In general, the stress tensor should characterize how the 
action responds to a diffeomorphism acting on the boundary, for which $\delta_\xi q_{ij} = \lie_\xi 
q_{ij}$ and $\delta_\xi n^i = \lie_\xi n^i$.  Using the expression (\ref{eqn:delSnull}) for 
a general variation of the action  and dropping terms that localize to the 
boundary of $\ns$, we find that $\delta_\xi S^\text{cl} = -\int_\ns \beom_\xi$, with
\beq\label{eqn:beomxi}
\beom_\xi = -\pi^{ij} \lie_\xi q_{ij} - \pi_i \lie_\xi n^i = -\pi^{ij} (\xi^k D_k q_{ij} + 2 D_i \xi^k
q_{kj}) - \pi_i (\xi^k D_k n^i - n^k D_k \xi^i),
\eeq
where $D_i$ is taken for the moment to be an arbitrary torsionless affine connection 
on $\ns$.  Since the first equality here only involves Lie derivatives, the expression
does not depend on the choice of $D_i$.
The expression on the right hand side of (\ref{eqn:beomxi}) can be rearranged
to express the equation in the form
\beq \label{eqn:fluxbalance}
\beom_\xi =  dj_\xi - f_\xi,
\eeq
where $j_\xi$ and $f_\xi$ are unambiguously determined by requiring that they
both depend
linearly and algebraically on $\xi^i$.  We will use this decomposition to define the 
stress tensor by the relation $j_\xi = -T\ind{^i_j} \xi^j \eta_i$
(see footnote \ref{ftn:orientation} regarding this choice of sign), 
as well as a generalized
divergence by the equation $f_\xi = -(\dv T)_j \xi^j\eta$.  Using the undensitized momenta
$(p^{ij}, p_i)$, defined by $\pi^{ij} = \eta p^{ij}$, $\pi_i = \eta p_i$, 
the boundary stress tensor and its generalized divergence are found to be\footnote{These expressions
for a stress tensor and generalized divergence are not special to general relativity, but instead
hold for any theory whose action is a functional of a Carroll structure, $S[q_{ij} ,n^i]$.  
The Carrollian momenta $p^{ij}$ and $p_i$ can be defined by  variational derivatives
of such an action with respect to $q_{ij}$ and $n^i$, and the stress tensor 
of the theory is still given by (\ref{eqn:nullTgen}).  This allows generalizations not
only to other theories of gravity possessing a Dirichlet variational principle,
but also to more general Carrollian field theories defined intrinsically on a null surface 
(see e.g.\ \cite{Bagchi:2019xfx}).
A similar expression for a stress tensor of asymptotically 
flat 3D gravity in terms of Carrollian momenta was presented in 
\cite{Hartong:2015usd}.  
A related construction of Carrollian momenta and conservation 
laws was considered in \cite{Ciambelli:2018ojf}, although they
utilize a slightly different set of geometric structures and momenta.  
Their conservation laws are naturally interpreted as a null limit of 
ordinary covariant conservation with respect to a pseudo-Riemannian connection,
as explored in \cite{Ciambelli:2018xat, Ciambelli:2018wre}.}
\begin{align}
T\ind{^i_j} &= 2 p^{ik}q_{kj} - n^i p_j \label{eqn:nullTgen}\\
(\dv T)_j &= D\ind{_i} T\ind{^i_j} - \rho_i T\ind{^i_j} - p^{ik}D_j q_{ik} - p_i D_j n^i
\label{eqn:divT}
\end{align}
Note that the corrections appearing in the generalized divergence are similar to those 
that occur in the divergence formula (\ref{eqn:nulldiv}) for a vector field with respect to a 
connection $D_i$ that does not preserve the volume form.
Although the expression for $(\dv T)_j$ appears to depend on the choice of connection,
such dependence is superficial, as can be seen by noting that 
\beq
(\dv T)_j\xi^j \eta =  d(T\ind{^i_j}\xi^j \eta_i)
-\pi^{ij}\lie_\xi q_{ij} - \pi_i \lie_\xi n^i,
\eeq
with all terms on the right hand side manifestly independent of the 
connection.\footnote{Such a generalized divergence can be defined for any tensor 
$A\ind{^i_j}$ for which $n^j$ is an eigenvector, $A\ind{^i_j}n^j = \alpha n^i$, and 
whose spatial component is symmetric, $A\ind{^i_j} q_{ik} = A\ind{^i_k} q_{ij}$. Any such tensor
can be decomposed as $A\ind{^i_j} = 2a^{ik}q_{kj} - n^i a_j$ with $a^{ij}$ symmetric, and 
the pair $(a^{ij} ,a_j)$ are only determined up to shifts of the form $a^{ij}\rightarrow a^{ij}
+V^i n^j + n^i V^j$, $a_j\rightarrow a_j + q_{jk}V^k$.  Defining the divergence by 
\beq
(\dv A)_j \xi^j \eta = d(A\ind{^i_j} \xi^j\eta_i) - \eta(a^{ij}\lie_\xi q_{ij} + a_i \lie_\xi n^i),
\eeq
one can check that the resulting expression is insensitive to the ambiguity in the 
definition of $a^{ij}$ and $a_i$.  
}
The expression (\ref{eqn:nullTgen}) gives the null version of the Brown-York stress tensor,
which can be rearranged using the expressions (\ref{eqn:piij}) and (\ref{eqn:pii})
and the decomposition (\ref{eqn:Wdecomp}) of the shape operator to give
\beq\label{eqn:nullBY}
T\ind{^i_j} = -\frac{1}{8\pi G}\left(W\ind{^i_j} - W \delta\ind{^i_j}\right)
\eeq
where $W = W\ind{^i_i} = \Theta + k$.  This expression is exactly analogous to the 
timelike Brown-York stress tensor (\ref{eqn:TBYdefn}), with the null shape operator 
$W\ind{^i_j}$ replacing the extrinsic curvature tensor $K\ind{^i_j}$ in the timelike
case.  In fact, $K\ind{^i_j}$ has the interpretation of a shape operator
for a timelike surface, which further tightens the analogy between the two cases.
An important property of the null Brown-York stress tensor (\ref{eqn:nullBY}) is 
that because it is constructed directly from $W\ind{^i_j}$, 
it is completely independent of the choice of $l_i$, despite a superficial
dependence on this choice in the expressions for the individual momenta
$p^{ij}$ and $p_i$.  It is worth pointing out that the mixed index structure $T\ind{^i_j}$
is important for obtaining an object that is independent of $l_i$:
any procedure for raising or lowering an index to obtain tensors $T^{ij}$ or $T_{ij}$ will
necessarily introduce dependence on some auxiliary structure such as $l_i$, or else kill some 
components if, for example, the degenerate metric $q_{ij}$ is used to lower an index.

To prove equality between the Brown-York current $j_\xi$
and the charge density $h_\xi$ constructed
using the Wald-Zoupas procedure \cite{Chandrasekaran:2020wwn},
we must show as in the timelike case that the divergence of the stress 
tensor (\ref{eqn:divT}) vanishes as a consequence of the constraint equations on $\ns$.  
To do so, we now take the connection $D_i$ to be 
an induced rigged connection, satisfying 
equations (\ref{eqn:Dq}-\ref{eqn:Dmu}).  
The first term in $\dv T$ takes the expected from as a covariant divergence
with respect to the connection $D_i$.  The remaining
terms can be shown to cancel:
\begin{align}
&2\rho_i W\ind{^i_j}- 2W\rho_j + \left(K^{ik} - W q^{ik}\right)
(l_i K_{jk} + l_k K_{ji})
- 2(\varpi_i + \Theta l_i) W\ind{^i_j}
\nonumber \\
&=\; -2\big[-\rho_i W\ind{^i_j}+ W\rho_j+ \rho_i W\ind{^i_j}
+(\Theta + k)l_i W\ind{^i_j}
\big] \nonumber\\
&=\; 0
\end{align}
where we have used that $l_i K^{ik} = l_i q^{ik} = 0$ and $\varpi_i = \rho_i + k l_i$.  Hence,
the divergence of the stress tensor when evaluated using an induced rigged connection
is given simply by
\beq
(\dv T)_j = D\ind{_i}T\ind{^i_j}.
\eeq

The final step is to relate this divergence to the constraint equations on $\ns$.  
The contracted Codazzi equation of the null surface
is given by equation (6.3) reference 
\cite{Gourgoulhon:2005ng}, which evaluates to
\begin{align}
\Pi\ind{^b_c}R_{ab} n^a &=\Pi\ind{^b_c}\left[ \nabla_a K\ind{^a_b} + n^a\nabla_a \rho_b +(k+\Theta)\rho_b
-\nabla_b(k+\Theta) -K_{bc}l^a\nabla_a n^c\right] \\
&=
D_a K\ind{^a_c}+ n^a D_a\rho_c +W \rho_c-D_cW \\
&=
D_a W\ind{^a_c} - D_c W
\end{align}
where to get to the second line we used $\nabla_a K\ind{^a_b} =\Pi\ind{^a_d}\nabla_a
K\ind{^d_b} -n_a l^c\nabla_c K\ind{^a_b}
= \Pi\ind{^a_d}\nabla_a K\ind{^d_b}+l^c\nabla_c n^a K_{ab}$, and for the third
line we used $D_a n^a = W$.  We immediately recognize this to be proportional to 
the divergence of the null Brown-York stress tensor, and hence we conclude that this divergence
vanishes on shell,
\beq \label{eqn:nullconservation}
D\ind{_i} T\ind{^i_j} = -8\pi G \Pi\ind{^b_j} R_{ab} n^a = 0.
\eeq
Often, the null Codazzi equations are separated into the $n^j$ component, which
is just the Raychaudhuri equation, and a spatial component, which is known as 
the Damour-Navier-Stokes equation \cite{Damour1982, Gourgoulhon:2005ng}.

As in the timelike case, the vanishing of $(\dv T)_j$ allows us to now conclude
the equality of the Wald-Zoupas and Brown-York charge densities.  The Wald-Zoupas charge
densities associated with a Dirichlet flux condition
are again given by the general formula (\ref{eqn:hxi}) using the 
the expressions (\ref{eqn:nullbdy}) and 
(\ref{eqn:nullcorner}) for $\ell$ and $\beta$.
They satisfy the continuity equation $d h_\xi = \beom_\xi$ 
\cite{Chandrasekaran:2020wwn}, which,
according to (\ref{eqn:beomxi}), is equal to $dj_\xi$ after imposing the stress
tensor conservation equation $f_\xi =0$.  
We can therefore conclude the equality of $h_\xi$ and $j_\xi$, subject to the 
same caveats described above equation (\ref{eqn:HxiQxiBY})
regarding the addition of exact forms $d c_\xi$ to each.  This 
then demonstrates that for symmetries that act covariantly on $q_{ij}$ and $n^i$,
\beq \label{eqn:HxiQxi}
H_\xi = \int_{\cal S} h_\xi = \int_{\cal S} j_\xi = \byc_\xi.
\eeq
The on-shell continuity equation $dh_\xi = dj_\xi = \beom_\xi$ implies that the 
Brown-York charges (\ref{eqn:HxiQxi}) are not conserved between cuts of the boundary 
$\ns$, but instead satisfy a flux-balance equation in which the difference of the 
charges is given by the integral of $\beom_\xi$ between the cuts \cite{Chandrasekaran:2020wwn,
Chandrasekaran2021}.

When utilizing additional intrinsic boundary terms during holographic renormalization
procedures for asymptotic charges, the two notions of charges will continue to agree,
provided the boundary counterterms $\ell_\text{ct}$ are fully covariant
with respect to the diffeomorphisms from which the charges are being constructed.

\subsection{Comparison to other null Brown-York stress tensors}
\label{sec:compareby}

While the expression (\ref{eqn:nullBY}) for the Brown-York stress tensor on a null
surface is a novel result of the present work, the idea of applying the Brown-York 
construction to null surfaces has been considered previously, see e.g.\
\cite{Brown:1996bw, Booth:2001gx, Jafari:2019bpw, Donnay2019}.  Of particular note are the works of 
Jafari \cite{Jafari:2019bpw} and Donnay and Marteau \cite{Donnay2019}, which both offer proposals for 
a full stress tensor associated with a null surface.  These proposals each differ slightly
from the stress tensor (\ref{eqn:nullBY}) due to different choices in boundary conditions
and intrinsic structures on the null surface when defining the subregion variational principle, 
and in this section we briefly describe the 
difference between these various proposals for the null Brown-York stress tensor.  
In making comparisons, we will refer to the stress tensor (\ref{eqn:nullBY}) of 
the present work as the 
{\it normal Brown-York stress tensor}, since it arises from a variational principle 
that fixes the null normal $n_a$ at $\ns$.  

The stress tensor defined by Jafari \cite{Jafari:2019bpw} bears many similarities to (\ref{eqn:nullBY}),
as both are derived from a Dirichlet variational principle.  The main difference is that 
Jafari does not impose the boundary condition that the surface $\ns$ remain null for all
variations of the metric, and the variations that change the null character of the 
surface are related to the energy density computed by the stress tensor.  Additionally,
Jafari employs a foliation by codimension-2 surfaces in the region of spacetime near
$\ns$, following the constructions 
in Refs. \cite{Hopfmuller2017a, Hopfmuller2018, Aghapour:2018icu},
which induces a preferred foliation of the null hypersurface.  This has the effect
of allowing for some variations that rescale the null generator, $\delta n_a \propto n_a$, in
contrast to the boundary condition $\delta n_a = 0$ employed in the present work.  A final
difference  is that Jafari obtains a stress tensor with covariant indices $T_{ij}$, making the 
comparison to the mixed index version $T\ind{^i_j}$ somewhat subtle.  However, 
because the null surface in Jafari's construction
comes equipped with a preferred auxiliary null vector $l^a$ due to the local foliation
by codimension-2 surfaces, a prescription can be given to define an equivalent mixed-index 
tensor.  This amounts to defining an intrinsic Lorentzian metric on the null surface
using the induced auxiliary one-form $l_i$ via
\beq
h_{ij} = -l_i l_j + q_{ij},
\eeq
whose inverse is given by $h^{ij} = -n^i n^j + q^{ij}$.  This choice then implies that 
lowering the index of the null generator $n^i$ yields the auxiliary one-form, $n^i h_{ij} = l_j$.
When computing the components of the stress tensor, this prescription can be implemented
through the relation $T\ind{^i_j} l_i= T_{ij} n^i$.

The most straightforward way to compare to Jafari's expressions is then to decompose the 
normal Brown-York
stress tensor (\ref{eqn:nullBY}) into an energy density $E$, momentum $P_i$,  and 
a spatial stress tensor $\Sigma\ind{^i_j}$ according to 
\begin{align}
E &= -T\ind{^i_j}n^j l_i = \frac{\Theta}{8\pi G} \label{eqn:BYE}\\
P_k&=  T\ind{^i_j} l_i q\ind{^j_k} = \frac{\varpi_k}{8\pi G} \label{eqn:BYP}\\
\Sigma\ind{^i_j} &= T\ind{^k_l} q\ind{^l_j} q\ind{^i_k} = -\frac{1}{8\pi G}\left(K\ind{^i_j} - 
(\Theta + k) q\ind{^i_j}\right). \label{eqn:BYstress}
\end{align}
The expressions for the momentum density $P_j$ and the spatial stress tensor $\Sigma\ind{^i_j}$
coincide with Jafari's expressions in equations (37) and (38) of 
Ref. \cite{Jafari:2019bpw}; however, 
the energy density (\ref{eqn:BYE}) differs from Jafari's expression, which instead involves
the expansion
of the codimension-2 foliation along the auxiliary null direction $l^a$ 
transverse to the surface.  The difference in the expression for the energy density 
is entirely due to the different choice of background structures and boundary conditions
Jafari employs in defining the null surface variational principle.

The stress tensor of Donnay and Marteau \cite{Donnay2019}
is obtained by a somewhat different procedure.
Rather than working directly on the null surface, they consider a sequence of timelike 
surfaces that limit to the null surface.  On each timelike surface, the usual Brown-York stress
tensor can be constructed from the extrinsic curvature $K\ind{^i_j}$
as in (\ref{eqn:TBYdefn}).  Although this diverges as the null limit is taken, the densitized
stress tensor $T\ind{^i_j}\eta_i$ has a finite limit, which defines a tensor $(T_\text{NL})\ind{^i_j}$
on $\ns$ that we refer to 
as the {\it null-limit Brown-York stress tensor}.  

To see how this works in more detail, it is helpful to consider an unnormalized normal vector
to the timelike hypersurfaces, which smoothly limits to the null normal $n_a$ of the null
surface.  Hence, we consider a function $\Phi$ whose level sets foliate the region near 
$\ns$, such that the unnormalized normal $n_a =\nabla_a\Phi$ limits to the null normal 
at $\Phi=0$. The norm of $n_a$ 
vanishes as the null surface is approached, and hence there must be a function 
$\kappa$ which we refer to as the  {\it surface gravity}, that satisfies the equation
\beq
g^{ab} n_a n_b = 2\kappa \Phi,
\label{surfacegravity}
\eeq
and generically $\kappa$ has a nonzero limit to $\ns$.\footnote{For generic null surfaces,
the surface gravity $\kappa$ defined by equation (\ref{surfacegravity}) can differ from the 
inaffinity $k$ defined by equation (\ref{eqn:inaffinity}), although these two
definitions agree for Killing horizons, as well as whenever $n_a$ is chosen
to be a pure gradient.}  In terms of $n_a$, the projector onto the timelike surfaces
can be expressed as 
\beq
h\ind{^a_b} = \delta\ind{^a_b} -\frac{n^a n_b}{2\kappa\Phi}.
\eeq
Using this projector, we can construct a shape operator 
$\nn K\ind{^a_b}$from the covariant derivative 
of the normal by the relation
\beq \label{eqn:Kunnorm}
\nn K\ind{^a_b} = h\ind{^a_c}h\ind{^d_b}\nabla_d n^c = h\ind{^d_b}\left[\nabla_d n^a -\frac12n^a\nabla_d
\log\kappa\right]
\eeq
where in obtaining the final expression on the right hand side of \eqref{eqn:Kunnorm}, we used the fact that $h^{a}{}_{b} \nabla_{a} \Phi = h^{a}{}_{b} n_{a}=0$. 
By definition, this tensor is tangential to the constant $\Phi$ surfaces on its contravariant
index, and hence pulls back to a well defined tensor $\nn K\ind{^i_j}$ on the surface,
whose defining relation is 
\beq\label{eqn:Kpb}
\nn K\ind{^i_j} \Pi\ind{^a_i} = \Pi\ind{^b_j} 
\left[\nabla_b n^a -\frac12n^a\nabla_b
\log\kappa\right]
\eeq  
where $\Pi\ind{^a_i}$ is the pullback map to each constant $\Phi$ surface.
It is related to the more familiar extrinsic curvature of the surface $K\ind{^i_j}$, which is 
constructed from the covariant derivative of the 
unit normal $\h n^a = \frac{n^a}{\sqrt{2\kappa \Phi}}$ by a simple
rescaling \cite{Mars1993},
\beq \label{eqn:Knorm}
K\ind{^i_j} = \frac{1}{\sqrt{2\kappa \Phi}} \nn K\ind{^i_j}.
\eeq
The terms appearing in (\ref{eqn:Kpb}) have manifestly finite limits to $\ns$, and we 
see that $\nn K\ind{^i_j}$ limits to a shifted version of the null surface shape operator,
\beq
\nn K\ind{^i_j} \overset{\Phi\rightarrow 0}\longrightarrow 
\:W\ind{^i_j} -n^i a_j
\eeq
where 
\beq
a_j = \frac12 D_j \log\kappa
\eeq
is the acceleration of the normal vector of the timelike foliation.

Since $\nn K\ind{^i_j}$ has a finite limit, it is clear from equation (\ref{eqn:Knorm}) that the 
extrinsic curvature tensor $K\ind{^i_j}$ diverges as $\Phi^{-\frac12}$ in the null
limit.  The Brown-York tensor on the timelike surfaces will then similarly diverge in the null
limit.  However, as discussed at the beginning of
section \ref{sec:nullbdy}, a more natural object to consider is the 
densitized stress tensor $T\ind{^i_j}\h\eta_i$, where $\h \eta$ is the induced 
volume form. Since $\h\eta$ vanishes as $\sqrt{2\kappa \Phi}$ as the surface becomes null,
we see that the densitized stress tensor has a finite null limit.  We can then turn this 
into a tensor on the null surface using the volume form $\eta$ associated with the null normal
$n_a$ to obtain the null-limit  Brown-York stress tensor,
\beq
(T_{\text{NL}})\ind{^i_j} = -\frac1{8\pi G}\left(\nn K\ind{^i_j} - \nn K\delta \ind{^i_j} \right)
=-\frac{1}{8\pi G}\left(W\ind{^i_j} - W \delta\ind{^i_j} -n^i a_j 
+(n^k a_k) \delta\ind{^i_j} \right).
\eeq
This final expression is the stress tensor obtained by Donnay and Marteau
\cite{Donnay2019}, 
which differs from the normal Brown-York stress tensor (\ref{eqn:nullBY}) by the acceleration 
terms $a_j$ which involve
 gradients of the surface gravity,
\beq
(T_\text{NL})\ind{^i_j} = T\ind{^i_j}  
+ \frac{1}{8\pi G}\left(n^i a_j - n^k a_k \delta\ind{^i_j}\right).
\eeq
An interesting feature is that these corrections cancel out of the energy density, which
is still given by (\ref{eqn:BYE}) when using the null-limit stress tensor.  
Gradients of $\kappa$ then enter into expressions for the momentum $P_k$ and 
the spatial stress tensor $\Sigma\ind{^i_j}$.

Finally, we mention that the conservation equation for the null-limit stress tensor
is somewhat different from the conservation equation (\ref{eqn:nullconservation})
for the normal Brown-York stress tensor
(\ref{eqn:nullBY}).
Continuing to employ the induced rigged connection defined  in \ref{eqn:DiVj}
and using that $D_i(W\ind{^i_j} - W\delta\ind{^i_j})=0$ on shell
and $da = 0$,
we find that 
\begin{align}
D_i (T_\text{NL})\ind{^i_j} &
= \frac{1}{8\pi G} D_i\left(
n^i a_j -\delta\ind{^i_j}(n^k a_k) \right) \nonumber \\
&= 
-\frac{1}{8\pi G} \left(W\ind{^i_j} - W \delta\ind{^i_j}\right)a_i
\nonumber \\
&=(T_\text{NL})\ind{^i_j}a_i
\label{eqn:DTNL}
\end{align}
where the last line uses $a_i (n^i a_j - \delta\ind{^i_j} n^k a_k) = 0$.
This correction to the conservation equation involving the acceleration is expected from
the perspective of the null limit, since $(T_\text{NL})\ind{^i_j}$ arises as the limit of 
rescaled version of the timelike Brown-York stress tensor,
\beq
(T_\text{NL})\ind{^i_j} = \lim_{\Phi\rightarrow 0} \left[-\frac{\sqrt{2\kappa \Phi}}{8\pi G}
\left( K\ind{^i_j} - K \delta\ind{^i_j}
\right) \right].
\eeq
Since the usual Brown-York stress tensor constructed from $K\ind{^i_j}$ is covariantly conserved 
on-shell
with respect to the induced connection on the timelike surfaces, the rescaled stress tensor
satisfies a modified conservation equation exactly of the form (\ref{eqn:DTNL}).
It would be interesting to explore in more detail these modifications of the conservation 
equation, and their relation to the Carrollian conservation 
equations described in \cite{Donnay2019, Ciambelli:2018xat, Ciambelli:2018wre, 
Ciambelli:2018ojf}.
In particular, there may be a different notion of connection on the null surface that 
is naturally induced from the null limit that makes the interpretation of the conservation
more straightforward.

\section{Anomalous transformations}
\label{sec:byanom}

The previous sections demonstrated that the Brown-York and
canonical Wald-Zoupas charges
agree, provided 
that the diffeomorphisms being considered act covariantly on the intrinsic
boundary structures and on the boundary term $\ell$ in the action.  
In some contexts, however, it is useful to consider charges for transformations
that act anomalously on these structures. 
For example, the appearance of central extensions in 
the algebra of 
Virasoro charges on black hole horizons 
\cite{Haco:2018ske, Chen:2020nyh}
was shown in \cite{Chandrasekaran:2020wwn} to be 
a consequence of such anomalous transformations.  
They arise in situations where additional background structure is introduced 
when constructing the subregion action that was not originally present in the 
definition of the field configuration space.
Transformations
which do not preserve the background structure will act anomalously on functionals
that depend on it.

In this section, we investigate whether the two notions of charges continue 
to agree when considering these more general, anomalous transformations.  For finite
timelike surfaces, all diffeomorphisms that are tangent to the surface act 
covariantly on the GHY boundary term (\ref{eqn:GHY}) and the intrinsic
metric $h_{ij}$, and hence the equality between Brown-York and canonical 
charges holds. 
 By contrast, it was shown in 
\cite{Chandrasekaran:2020wwn} that certain diffeomorphisms of a null surface can act
anomalously on the boundary 
term (\ref{eqn:nullbdy})  and intrinsic quantities.  For this reason,
we restrict attention to null surfaces in this section.  In principle, 
anomalous transformations
can also arise for timelike surfaces if the diffeomorphism contains a nonzero 
transverse component.  This situation is particularly relevant for asymptotic
symmetries of anti-de Sitter space, where such transformations
give rise to the well-known holographic Weyl anomalies of the dual CFT
\cite{Witten:1998qj,Henningson1998a, Balasubramanian:1999re, DeHaro2001,
Papadimitriou:2005ii}.  
Although we do not analyze this case in detail due to 
subtleties in handling transverse surface deformations, we expect similar 
reasoning to apply in this case as well.  It would be interesting to 
explore this case in more detail in the future.

As discussed in section \ref{sec:nullby}, since the Dirichlet variational principle
with a null boundary is formulated by fixing a preferred null normal $n_a$, 
anomalies can arise from diffeomorphisms which are tangent to $\ns$ but only preserve
$n_a$ up to a rescaling.  This rescaling $w_\xi$ is known as the boost weight
of the transformation, and is defined through the equation
\beq
\lie_\xi n_a = w_\xi n_a.
\eeq
On the other hand, since $n_a$ is taken to be a fixed quantity, it cannot
in particular transform nontrivially under diffeomorphisms, so that 
$\delta_\xi n_a =0$.  It is convenient to parameterize the failure of 
$n_a$ to transform covariantly through an anomaly operator $\Delta_\xi$ 
\cite{Hopfmuller2018}, defined by 
\beq
\delta_\xi n_a = \lie_\xi n_a + \Delta_\xi n_a.
\eeq
By the above discussion, we immediately find $\Delta_\xi n_a = -w_\xi n_a$, 
and this equation also fixes the anomalous transformation of the intrinsic
null generator,
\beq \label{eqn:Delxini}
\Delta_\xi n^i = - w_\xi n^i.
\eeq
Such transformations also act anomalously on the null boundary term
(\ref{eqn:nullbdy}), and this anomaly was computed in \cite{Chandrasekaran:2020wwn}
to be
\beq \label{eqn:Delxiell}
\Delta_\xi \ell = \frac{(n^a\nabla_a w_\xi) \eta}{8\pi G}.
\eeq
Since the degenerate induced metric is independent of 
the normalization of $n_a$, it continues to transform covariantly,
$\Delta_\xi q_{ij} = 0$.

Note that the conservation equation (\ref{eqn:nullconservation}) for the null Brown-York 
stress tensor is independent of the vector field $\xi^a$ and whether 
or not it acts anomalously.  Because of this, once
the gravitational constraint equations are imposed, 
the divergence of the 
Brown-York charge density always satisfies
\beq
dj_\xi = \beom_\xi,
\eeq
with $\beom_\xi$ given in (\ref{eqn:beomxi}).  On the other hand,
the canonical charge density $h_\xi$ satisfies a modified continuity equation
derived in \cite{Chandrasekaran:2020wwn}, 
\begin{align}
dh_\xi &= -\pi^{ij}\lie_\xi q_{ij} - \pi_i \lie_\xi n^i - \pi_i \Delta_\h\xi n^i
-\Delta_\h\xi \ell \\
&= dj_\xi + (\dv T)_j \xi^j\eta -\frac{\eta}{8\pi G}( \Theta w_\xi + n^iD_i w_\xi).
\label{eqn:anomdhxi}
\end{align}
The final term in (\ref{eqn:anomdhxi})
reduces to the exact term $-\frac{1}{8\pi G} d(w_\xi \mu)$ on the null surface.  
We can therefore conclude that on shell,
the Brown-York and Wald-Zoupas   charge densities are related via
\beq
h_\xi = j_\xi - \frac{1}{8\pi G}w_\xi \mu,
\eeq
and therefore the charges differ by the integral of the anomaly
over the cut $\cal S$,
\beq\label{eqn:HvsBY}
H_\xi = \byc_\xi -\frac{1}{8\pi G}\int_{\mathcal{S}} w_\xi\mu.
\eeq

As an explicit 
example of how such a difference between Brown-York and Wald-Zoupas charges
can arise, we now consider
the symmetries and charges at a finite null boundary found by 
Chandrasekaran, Flanagan, and Prabhu (CFP) \cite{CFP}.\footnote{The same 
algebra was recently considered as an extended symmetry algebra 
of null infinity in Ref. \cite{freidel2021weyl}.} 
In this example, the field configuration space for general relativity 
in a subregion with a null boundary is further restricted to fix
$n^i$ and $k$ on the null surface.  
It was shown that the symmetry group of this field configuration space is 
\begin{align}
    \text{diff}(\mathcal{S})\ltimes \mathfrak{s}, 
\end{align}
where $\mathcal{S}$ is the base space of $\mathcal{N}$, viewing $\mathcal{N}$ as a fiber bundle where the null generators are fibered over $\mathcal{S}$,
and $\mathfrak{s}$ consists of the set of vector fields $\xi^a = f n^a$ for functions $f:\mathcal{S} \rightarrow \mathbb{R}$ which satisfy 
\begin{align}
    \lie_n (\lie_n + \kappa)f = 0. \label{supertranslation}
\end{align}
The solutions correspond to angle-dependent translations and angle-dependent rescalings of the integral curves of the null generator. The $\text{diff}(\mathcal{S})$ generators are represented by vector fields $\xi^a = X^a$ where $X^a n_a = X^a l_a = 0$.  
The Wald-Zoupas charges associated to this symmetry group were computed in CFP
using covariant phase space methods, with the result that on any cross-section $S$ of the null boundary,\footnote{In CFP the Wald-Zoupas charges were found using the stationarity condition that the flux should vanish for all perturbations on a solution for which $\mathcal{N}$ is stationary. For the field space considered in CFP,
which imposes boundary conditions fixing $n^i$ and $k$, 
this happens to agree with the Dirichlet variational principle adopted in the present paper.}
\begin{align}
    H_{\xi} = -\frac{1}{8\pi G}\int_{\mathcal{S}} \mu~\left(\Theta f - \lie_n f + W^{i}{}{}_j X^j l_i - l_i \lie_X n^i \right), \label{wzc}
\end{align}
where we have shifted the boundary term used in CFP  by 
$\ell \rightarrow \ell - \frac{1}{8\pi G}( k- \Theta) \eta$
in order to agree with the choice of boundary term and total derivative term $\beta$
in the present work. 
We now compare these charges to the ones which result from the Brown-York stress tensor \eqref{eqn:nullBY}. 

The Brown-York charges for these symmetries are
\begin{align}
    \byc_\xi = -\int_\mathcal{S} T\ind{^i_j} \xi^j \eta_i
    =-\frac{1}{8\pi G}\int_{\mathcal{S}} \mu~\left(\Theta f + W^i{}{}_j X^j l_i\right),  \label{byc}
\end{align}
which differs from the Wald-Zoupas charges. However, note that the field space defined in CFP contains an 
anomalous transformation of $n^i$ (\ref{eqn:Delxini}).
A simple computation of $w_\xi$ results in 
\begin{align}
    w_{\xi} = -\lie_n f - l_i \lie_X n^i. 
\end{align}
Therefore, the Wald-Zoupas and Brown-York charges differ by the term 
\begin{align}\label{eqn:H-Q}
  H_{\xi} - \byc_\xi = -\frac{1}{8 \pi G}\int_{\mathcal{S}}\mu ~w_{\xi},
\end{align}
in agreement with the general relation \eqref{eqn:HvsBY}.

We further remark on an important difference between the Wald-Zoupas and Brown-York charges. Choose coordinates $(u, x^A)$ on $\mathcal{N}$, where $u$ is an affine parameter for the null generator, i.e.\ $n^i = \partial_u,~ k =0$, and $x^A$ are coordinates for $\mathcal{S}$. Furthermore, we can always choose $u$ such that $\mathcal{S}$ is at $u = 0$. The condition \eqref{supertranslation} can be solved to get $f(u, x^A) = f_0(x^A) + u f_1(x^A)$. The angle-dependent translations, $f_0(x^A)$, contribute to both \eqref{wzc} and \eqref{byc}, but the angle-dependent rescalings $f_1(x^A)$ only contribute to \eqref{wzc}. In particular, they only contribute through the anomaly term. Ultimately this is because the rescaling generators come from boost generators in the bulk spacetime, which vanish at $\mathcal{S}$ but have non-vanishing first derivative there. While the 
Wald-Zoupas charges depend on first derivatives of the symmetry generators, the Brown-York charges do not.\footnote{Of course, if we evaluate the Brown-York charges for the rescaling symmetries on any other cross-section, after having fixed $\mathcal{S}$ to lie at $u = 0$, then it will be non-vanishing.} Note that if we take $\mathcal{N}$ to be the Killing horizon of a stationary black hole, then the Wald entropy \cite{Wald:1993nt} arises purely from the anomaly.

\section{Discussion and future work} \label{sec:discussion}

This paper has presented a novel expression for a Brown-York stress tensor associated 
with null hypersurfaces in general relativity and has established its two key features:
covariant conservation with respect to an induced connection on the null surface, and the 
relationship between the associated Brown-York charges and canonical charges obtained via the 
Wald-Zoupas construction.  The conservation equation was shown to be equivalent to 
the gravitational constraint equations on the null surface, consisting of the Raychaudhuri
and Damour-Navier-Stokes equations.  
The latter name refers to the analogy made by Damour between this evolution equation
for kinematical quantities on a black hole horizon and the Navier-Stokes equation of 
 hydrodynamics.  Reformulating it as a conservation equation for a stress tensor
sheds light on the reason for this analogy: 
at their core, hydrodynamical equations are simply conservation equations for a
fluid stress tensor.  The striking feature of the conservation equation (\ref{eqn:nullconservation})
for the null Brown-York stress tensor (\ref{eqn:nullBY}) is that it explicitly 
takes the form of a covariant divergence of stress tensor with respect to a well-defined
connection $D_i$  on the null surface.  This is in contrast to previous works
\cite{Damour1982, Donnay2019}, which generally separate these conservation equations
into components,  somewhat obscuring their interpretation.  It is also in contrast to 
formulations which express the conservation equations as flux balance equations such 
as in \cite{Hopfmuller2018}, which are more analogous to the on-shell 
relation $dj_\xi = \beom_\xi$ as in equation (\ref{eqn:fluxbalance}).
While imposing the flux balance equation for every choice of $\xi^i$ implies the conservation equation
and vice-versa, the conservation equation (\ref{eqn:nullconservation})
is more directly related to the dynamics of the 
system and the hydrodynamical equations of motion.

Note that the connection between
gravitational equations and hydrodynamics features prominently in the fluid-gravity correspondence
in holography \cite{Bhattacharyya:2007vjd, Rangamani:2009xk}.  
We therefore  expect the identification of the 
null Brown-York stress tensor and its conservation equation to yield important insights into
holographic correspondences involving null boundaries, including celestial holography 
and for holographic descriptions of subregions bounded by null surfaces.
Some ideas in this direction have been explored in \cite{Hartong:2015usd, Penna:2017vms,
Ciambelli:2018wre, Ciambelli:2020eba, Fareghbal:2013ifa}.
It would also be interesting to relate this Brown-York stress tensor with the stress tensor of celestial conformal field theory \cite{Kapec:2016jld} to see if there exists an equivalence between the two, analogous to that in AdS/CFT. This would be an important step in establishing
the holographic dictionary between general relativity in asymptotically flat spacetimes and celestial conformal field theory.

An immediate future direction would be to carry out this analysis at future null infinity in four-dimensional asymptotically flat spacetimes  to compare the Brown-York charges with the charges obtained, for example, in \cite{FN,GPS,Compere:2018ylh} for the symmetries corresponding to the (generalized) Bondi-Metzner-Sachs (BMS) algebra. 
For ordinary BMS symmetries in Bondi coordinates, it has been demonstrated in \cite{Brown:1996bw}
that the Brown-York charges reproduce the expression for BMS charges when
realizing null infinity as a null limit of timelike surfaces.  
More generally, however, applying the Brown-York construction to asymptotic
boundaries requires additional boundary counterterms $\ell^\text{ct}$ in
the subregion action to yield a finite renormalized on-shell action functional $S^\text{cl}$.
These counterterms must be suitably covariant to preserve the conservation equation 
of the stress tensor and to yield agreement with the canonical charges, and it is an interesting
question whether a fully covariant renormalized action can be obtained.  There are also choices in 
how one foliates spacetime near null infinity, and for some choices of foliation, the null-limit
stress tensor of Donnay and Marteau \cite{Donnay2019} may be a more appropriate object to consider, especially when
looking for covariant counterterms.  
It would further be interesting to determine how the continuity equation for the Brown-York
charges relates to the flux-balance equations for generalized BMS charges explored in
\cite{Compere:2019gft}; given that both are consequences of the Einstein equation, establishing 
a relationship should be possible.

Although we focused on vacuum general relativity in this work, generalizations involving 
the inclusion of matter or modified gravitational theories are possible.  As long as the 
modified action admits a Dirichlet variational principle, the general construction of the 
boundary stress tensor follows.  With matter fields, the on-shell action will  
also depend on the boundary values of the fields, and there will be contributions to 
the stress tensor coming from the matter action.  Similarly, higher curvature theories
such as Lovelock gravity will also yield a well-defined boundary stress tensor
utilizing  modified expressions for $\pi^{ij}$ and $\pi_i$ on the null surface
after adding the appropriate null boundary terms for these theories 
\cite{Chakraborty2019}.

\section*{Acknowledgments}
We thank Luca Ciambelli, Daniel Harlow, Rob Leigh, and Don Marolf 
for helpful discussions.  We also thank an anonymous
referee for comments that improved the presentation of this work.  E.F. and I.S. are supported in part by NSF
grants PHY-1707800 and PHY-2110463. I.S. also acknowledges support from the John and David Boochever prize fellowship in fundamental theoretical physics.
AJS is supported by the Air Force Office of Scientific Research under award number FA9550-19-1-036.
Research at Perimeter Institute is supported in part by the Government of Canada through the Department of Innovation, Science and Economic Development and by the Province of Ontario through the Ministry of Colleges and Universities. V.C. is supported in part by the Berkeley Center for Theoretical Physics; by the Department of Energy, Office of Science, Office
of High Energy Physics under QuantISED Award DE-SC0019380 and under contract DEAC02-05CH11231; by the National Science Foundation under grant PHY-1820912; and by a grant from the Simons Foundation (816048, VC).

\appendix
\section{Connections for Carroll geometries} \label{app:carroll}

In this appendix, we derive the compatibility relations for a class of connections associated 
with a Carroll geometry arising from a spacetime rigging vector.  As explained in 
section \ref{sec:geonullsurfaces},
the first step in defining an affine connection associated with a Carroll geometry $(q_{ij}, n^i)$
is to choose an Ehresmann connection, defined in terms of a one-form $l_i$, 
whose only requirement is that $l_i n^i = -1$.  This 
connection allows one to specify a class of horizontal vectors, consisting of any vector 
satisfying $V^i l_i = 0$.  The choice of $l_i$ already determines a set of differential invariants
coming from the exterior 
derivative $dl$
that are defined independently of any choice of affine connection.  
In general, this exterior derivative can be decomposed as 
\beq \label{eqn:dl}
dl = b \wedge l + r
\eeq
where $b_i$ and $r_{ij}$ are both horizontal differential forms.  We refer to $b_i$ as the boost
form and $r_{ij}$ as the curvature of the Ehresmann connection.\footnote{In
\cite{Ciambelli2019b}, the quantity $b_i$ was referred to as the ``acceleration'' and 
$r_{ij}$ as the ``Carrollian torsion.''} 

We wish to show that a connection $D_i$ on a Carroll manifold induced 
as a rigged connection associated with an embedded null surface 
in spacetime implies the relations (\ref{eqn:Dq}-\ref{eqn:Dmu}).  Additionally, we will
derive the expression for $D_i l_j$ arising from such a connection.  Equation (\ref{eqn:Dn})
follows immediately from the definitions of the induced connection and the shape operator,
(\ref{eqn:DiVj}) and (\ref{eqn:Wdefn}).  To derive (\ref{eqn:Dq}), we note that $q_{ij}$ 
maps via the inclusion $\Pi\ind{^i_a}$ to
a spacetime tensor $q_{ab}$ that annihilates both $n^a$ and $l^a$, given by
\beq
q_{ab} = g_{ab} + l_a n_b + n_a l_b.
\eeq
Then we can compute
\begin{align}
D_i q_{jk} &= \Pi\ind{^a_i}\Pi\ind{^b_j}\Pi\ind{^c_k}\left(\nabla_a(g_{bc}+l_b n_c + n_b l_c)\right)
\nonumber \\
&= \Pi\ind{^a_i}\Pi\ind{^b_j}\Pi\ind{^c_k}\left(l_b\nabla_a n_c + l_c \nabla_a n_b\right) 
\nonumber \\
&= l_j K_{ik} + l_k K_{ij}.
\end{align}

For the relation (\ref{eqn:Deta}), we note that the inclusion map sends $\eta$ defined intrinsically
on the null surface $\ns$ to a spacetime $(d-1)$-form $\eta_{a\ldots}$ 
that is related to the spacetime volume form
by the equation $\ep = n \wedge \eta$.  Furthermore $\eta_{a\ldots}$ must satisfy $l^a\eta_{a\ldots}=0$,
and hence it can be defined by the equation $\eta = -i_l \ep$.  Then we have 
\begin{align}
D_i \eta_{j\ldots} &= -\Pi\ind{^a_i} \Pi\ind{^b_j}\cdots \nabla_a l^c \ep_{cb\ldots}
\nonumber \\
&=-\Pi\ind{^a_i} \Pi\ind{^b_j}\cdots \nabla_a l^c \left(n_c \eta_{b\ldots} - (d-1) \eta_{c[\ldots} n_{b]}
\right) \nonumber \\
&= -\rho_i \eta_{j\ldots}.
\end{align}
Then using that $\mu = i_n \eta$ in $\ns$, relations (\ref{eqn:Dn}) and (\ref{eqn:Deta})
fix the expression for $D_i \mu$:
\begin{align}
D_i \mu &= D_i (n^j \eta_j) = W\ind{^j_i} \eta_j - n^j\rho_i \eta_j = K\ind{^j_i}\eta_j.
\end{align}

Finally, we should examine the expression for $D_i l_j$.  Since $D_i$ is torsionless,
the antisymmetric components are fixed in terms of $b_i$ and $r_{ij}$ in terms 
of (\ref{eqn:dl}).  
We can also compute the components parallel to $l_i$ by the relations
\begin{align}
n^j D_i l_j &= \Pi\ind{^a_i} n^b\nabla_a l_b = \rho_i \\
n^i D_i l_j &= \Pi\ind{^b_j} n^a \nabla_a  l_b = \Pi\ind{^b_j}(n^a(dl)_{ab} + n^a\nabla_b l_a)
= b_j + \rho_j.
\end{align}
The remaining components of $D_i l_j$ are purely horizontal, and the antisymmetric horizontal
component is simply $r_{ij}$ according to equation (\ref{eqn:dl}).  The remaining 
horizontal symmetric component is an independent tensor $\nu_{ij}$
characterizing the extrinsic geometry of the null surface.  Hence,
using $\rho_i = -k l_i + \varpi _i$, we arrive at the final decomposition
of $D_i l_j$,
\beq\label{eqn:Dl}
D_i l_j = k l_i l_j -\varpi_i l_j -l_i \varpi_j - l_i b_j + \frac12r_{ij} + \nu_{ij}.
\eeq

From the intrinsic perspective, we can ask to what extent these relations 
fully specify an affine connection that is in a certain sense compatible with the 
Carroll structure.  According to the above relations, we see that the additional data
needed to fix a torsionless connection on the intrinsic Carroll geometry is 
an Ehresmann connection $l_i$, a rotation one-form $\rho_i$, and a horizontal symmetric tensor
$\nu_{ij}$.  Choosing a coordinate system $(u, x^A)$ in which $n^i  = \partial_u^i$ and $x^A$ are 
coordinates for the horizontal directions, a generic Ehresmann connection can be parameterized
by $l_i = -\nabla_i u + w_B\nabla_i x^B$. 
Then, defining the connection coefficients in this coordinate system according to
$D_i V^j = \partial_i V^j +\gamma^i_{jk}V^k$, one can then check that 
equations (\ref{eqn:Dq}), (\ref{eqn:Dn}), and (\ref{eqn:Dl}) fully fix all 
components of $\gamma\ind{^i_{jk}}$, and hence these equations uniquely specify the connection
$D_i$.

Note that much of the work on connections associated with Carrollian structures has focused 
on connections that preserve $q_{ij}$ and $n^i$
\cite{Hartong:2015usd, Bekaert:2015xua, Ciambelli2019b}.  
When $K_{ij}$ is nonvanishing, these connections
necessarily must involve torsion \cite{Vogel1965}; for example, torsion is necessary
in order to be compatible with the equation $K_{ij} = \frac12 \lie_n
q_{ij}$ . 
In the present work, however, we found that the natural connection that arises in
considerations of the null Brown-York stress tensor and conservation equations involves a 
torsionless connection that generically does not preserve $q_{ij}$ and $n^i$.  Such connections
do not appear to have been considered in recent works on Carrollian geometry, 
and it would be interesting to further investigate the geometric properties of these connections 
in the future.
A related set of connections satisfying equation (\ref{eqn:Dq}) but not 
(\ref{eqn:Dn}) have been examined previously in 
\cite{Dautcourt1967}, and 
general properties of affine connections in Carrollian geometries have been
explored in \cite{Leigh-notes}.

\bibliographystyle{JHEPthesis}
\bibliography{BYcharges}

\providecommand{\href}[2]{#2}\begingroup\RaggedRight\begin{thebibliography}{10}

\bibitem{Barnich2010}
G.~Barnich and C.~Troessaert, \emph{{Aspects of the BMS/CFT correspondence}},
  \href{http://dx.doi.org/10.1007/JHEP05(2010)062}{\emph{J. High Energy Phys.}
  {\bf 05} (2010) 062}.
\newblock
  [\href{https://arxiv.org/abs/1001.1541}{\nolinkurl{arXiv:1001.1541}}].

\bibitem{Barnich:2011mi}
G.~Barnich and C.~Troessaert, \emph{{BMS charge algebra}},
  \href{http://dx.doi.org/10.1007/JHEP12(2011)105}{\emph{JHEP} {\bf 12} (2011)
  105}.
\newblock
  [\href{https://arxiv.org/abs/1106.0213}{\nolinkurl{arXiv:1106.0213}}].

\bibitem{Strominger:2013jfa}
A.~Strominger, \emph{{On BMS Invariance of Gravitational Scattering}},
  \href{http://dx.doi.org/10.1007/JHEP07(2014)152}{\emph{JHEP} {\bf 07} (2014)
  152}.
\newblock
  [\href{https://arxiv.org/abs/1312.2229}{\nolinkurl{arXiv:1312.2229}}].

\bibitem{He:2014laa}
T.~He, V.~Lysov, P.~Mitra and A.~Strominger, \emph{{BMS supertranslations and
  Weinberg\textquoteright{}s soft graviton theorem}},
  \href{http://dx.doi.org/10.1007/JHEP05(2015)151}{\emph{JHEP} {\bf 05} (2015)
  151}.
\newblock
  [\href{https://arxiv.org/abs/1401.7026}{\nolinkurl{arXiv:1401.7026}}].

\bibitem{Stro-lectures}
A.~Strominger, \emph{{Lectures on the Infrared Structure of Gravity and Gauge
  Theory}},
  [\href{https://arxiv.org/abs/1703.05448}{\nolinkurl{arXiv:1703.05448}}].

\bibitem{Campiglia:2014yka}
M.~Campiglia and A.~Laddha, \emph{{Asymptotic symmetries and subleading soft
  graviton theorem}},
  \href{http://dx.doi.org/10.1103/PhysRevD.90.124028}{\emph{Phys. Rev.} {\bf
  D90} (2014) 124028}.
\newblock
  [\href{https://arxiv.org/abs/1408.2228}{\nolinkurl{arXiv:1408.2228}}].

\bibitem{Campiglia:2015yka}
M.~Campiglia and A.~Laddha, \emph{{New symmetries for the Gravitational
  S-matrix}}, \href{http://dx.doi.org/10.1007/JHEP04(2015)076}{\emph{JHEP} {\bf
  04} (2015) 076}.
\newblock
  [\href{https://arxiv.org/abs/1502.02318}{\nolinkurl{arXiv:1502.02318}}].

\bibitem{Compere:2018ylh}
G.~Compere, A.~Fiorucci and R.~Ruzziconi, \emph{{Superboost transitions,
  refraction memory and super-Lorentz charge algebra}},
  \href{http://dx.doi.org/10.1007/JHEP11(2018)200}{\emph{JHEP} {\bf 11} (2018)
  200}.
\newblock
  [\href{https://arxiv.org/abs/1810.00377}{\nolinkurl{arXiv:1810.00377}}].

\bibitem{Flanagan:2019vbl}
E.~E. Flanagan, K.~Prabhu and I.~Shehzad, \emph{{Extensions of the asymptotic
  symmetry algebra of general relativity}},
  \href{http://dx.doi.org/10.1007/JHEP01(2020)002}{\emph{JHEP} {\bf 01} (2020)
  002}.
\newblock
  [\href{https://arxiv.org/abs/1910.04557}{\nolinkurl{arXiv:1910.04557}}].

\bibitem{freidel2021weyl}
L.~Freidel, R.~Oliveri, D.~Pranzetti and S.~Speziale, \emph{{The Weyl BMS group
  and Einstein's equations}},
  [\href{https://arxiv.org/abs/2104.05793}{\nolinkurl{arXiv:2104.05793}}].

\bibitem{Hollands:2005ya}
S.~Hollands, A.~Ishibashi and D.~Marolf, \emph{{Counter-term charges generate
  bulk symmetries}},
  \href{http://dx.doi.org/10.1103/PhysRevD.72.104025}{\emph{Phys. Rev. D} {\bf
  72} (2005) 104025}.
\newblock
  [\href{https://arxiv.org/abs/hep-th/0503105}{\nolinkurl{arXiv:hep-th/0503105}}].

\bibitem{Papadimitriou:2005ii}
I.~Papadimitriou and K.~Skenderis, \emph{{Thermodynamics of asymptotically
  locally AdS spacetimes}},
  \href{http://dx.doi.org/10.1088/1126-6708/2005/08/004}{\emph{JHEP} {\bf 08}
  (2005) 004}.
\newblock
  [\href{https://arxiv.org/abs/hep-th/0505190}{\nolinkurl{arXiv:hep-th/0505190}}].

\bibitem{Harlow:2019yfa}
D.~Harlow and J.-Q. Wu, \emph{{Covariant phase space with boundaries}},
  \href{http://dx.doi.org/10.1007/JHEP10(2020)146}{\emph{JHEP} {\bf 10} (2020)
  146}.
\newblock
  [\href{https://arxiv.org/abs/1906.08616}{\nolinkurl{arXiv:1906.08616}}].

\bibitem{Carlip_1999}
S.~Carlip, \emph{Black hole entropy from conformal field theory in any
  dimension}, \href{http://dx.doi.org/10.1103/physrevlett.82.2828}{\emph{Phys.
  Rev. Lett.} {\bf 82} (1999) 2828–2831}.
\newblock
  [\href{https://arxiv.org/abs/hep-th/9812013}{\nolinkurl{arXiv:hep-th/9812013}}].

\bibitem{Hawking_2016}
S.~W. Hawking, M.~J. Perry and A.~Strominger, \emph{Soft hair on black holes},
  \href{http://dx.doi.org/10.1103/physrevlett.116.231301}{\emph{Phys. Rev.
  Lett.} {\bf 116} (2016) }.
\newblock
  [\href{https://arxiv.org/abs/1601.00921}{\nolinkurl{arXiv:1601.00921}}].

\bibitem{Haco:2018ske}
S.~Haco, S.~W. Hawking, M.~J. Perry and A.~Strominger, \emph{{Black Hole
  Entropy and Soft Hair}},
  \href{http://dx.doi.org/10.1007/JHEP12(2018)098}{\emph{JHEP} {\bf 12} (2018)
  098}.
\newblock
  [\href{https://arxiv.org/abs/1810.01847}{\nolinkurl{arXiv:1810.01847}}].

\bibitem{Chen:2020nyh}
L.-Q. Chen, W.~Z. Chua, S.~Liu, A.~J. Speranza and B.~d. S.~L. Torres,
  \emph{{Virasoro hair and entropy for axisymmetric Killing horizons}},
  \href{http://dx.doi.org/10.1103/PhysRevLett.125.241302}{\emph{Phys. Rev.
  Lett.} {\bf 125} (2020) 241302}.
\newblock
  [\href{https://arxiv.org/abs/2006.02430}{\nolinkurl{arXiv:2006.02430}}].

\bibitem{Chandrasekaran:2020wwn}
V.~Chandrasekaran and A.~J. Speranza, \emph{{Anomalies in gravitational charge
  algebras of null boundaries and black hole entropy}},
  \href{http://dx.doi.org/10.1007/JHEP01(2021)137}{\emph{JHEP} {\bf 01} (2021)
  137}.
\newblock
  [\href{https://arxiv.org/abs/2009.10739}{\nolinkurl{arXiv:2009.10739}}].

\bibitem{Flanagan:2021ojq}
E.~E. Flanagan, \emph{{Order-Unity Correction to Hawking Radiation}},
  \href{http://dx.doi.org/10.1103/PhysRevLett.127.041301}{\emph{Phys. Rev.
  Lett.} {\bf 127} (2021) 041301}.
\newblock
  [\href{https://arxiv.org/abs/2102.04930}{\nolinkurl{arXiv:2102.04930}}].

\bibitem{Flanagan:2021svq}
E.~E. Flanagan, \emph{{Infrared effects in the late stages of black hole
  evaporation}}, \href{http://dx.doi.org/10.1007/JHEP07(2021)137}{\emph{JHEP}
  {\bf 07} (2021) 137}.
\newblock
  [\href{https://arxiv.org/abs/2102.13629}{\nolinkurl{arXiv:2102.13629}}].

\bibitem{Pasterski:2020xvn}
S.~Pasterski and H.~Verlinde, \emph{{HPS meets AMPS: How Soft Hair Dissolves
  the Firewall}},
  [\href{https://arxiv.org/abs/2012.03850}{\nolinkurl{arXiv:2012.03850}}].

\bibitem{Donnelly2016a}
W.~Donnelly and L.~Freidel, \emph{{Local subsystems in gauge theory and
  gravity}}, \href{http://dx.doi.org/10.1007/JHEP09(2016)102}{\emph{J. High
  Energy Phys.} {\bf 2016} (2016) 1--45}.
\newblock
  [\href{https://arxiv.org/abs/1601.04744}{\nolinkurl{arXiv:1601.04744}}].

\bibitem{Geiller:2017xad}
M.~Geiller, \emph{{Edge modes and corner ambiguities in 3d
  Chern\textendash{}Simons theory and gravity}},
  \href{http://dx.doi.org/10.1016/j.nuclphysb.2017.09.010}{\emph{Nucl. Phys. B}
  {\bf 924} (2017) 312--365}.
\newblock
  [\href{https://arxiv.org/abs/1703.04748}{\nolinkurl{arXiv:1703.04748}}].

\bibitem{Speranza2018a}
A.~J. Speranza, \emph{{Local phase space and edge modes for
  diffeomorphism-invariant theories}},
  \href{http://dx.doi.org/10.1007/JHEP02(2018)021}{\emph{J. High Energy Phys.}
  {\bf 2018} (2018) 1--37}.
\newblock
  [\href{https://arxiv.org/abs/1706.05061}{\nolinkurl{arXiv:1706.05061}}].

\bibitem{Geiller:2017whh}
M.~Geiller, \emph{{Lorentz-diffeomorphism edge modes in 3d gravity}},
  \href{http://dx.doi.org/10.1007/JHEP02(2018)029}{\emph{JHEP} {\bf 02} (2018)
  029}.
\newblock
  [\href{https://arxiv.org/abs/1712.05269}{\nolinkurl{arXiv:1712.05269}}].

\bibitem{Geiller:2019bti}
M.~Geiller and P.~Jai-akson, \emph{{Extended actions, dynamics of edge modes,
  and entanglement entropy}},
  \href{http://dx.doi.org/10.1007/JHEP09(2020)134}{\emph{JHEP} {\bf 09} (2020)
  134}.
\newblock
  [\href{https://arxiv.org/abs/1912.06025}{\nolinkurl{arXiv:1912.06025}}].

\bibitem{Freidel:2020xyx}
L.~Freidel, M.~Geiller and D.~Pranzetti, \emph{{Edge modes of gravity. Part I.
  Corner potentials and charges}},
  \href{http://dx.doi.org/10.1007/JHEP11(2020)026}{\emph{JHEP} {\bf 11} (2020)
  026}.
\newblock
  [\href{https://arxiv.org/abs/2006.12527}{\nolinkurl{arXiv:2006.12527}}].

\bibitem{Donnelly:2020xgu}
W.~Donnelly, L.~Freidel, S.~F. Moosavian and A.~J. Speranza,
  \emph{{Gravitational edge modes, coadjoint orbits, and hydrodynamics}},
  \href{http://dx.doi.org/10.1007/JHEP09(2021)008}{\emph{JHEP} {\bf 09} (2021)
  008}.
\newblock
  [\href{https://arxiv.org/abs/2012.10367}{\nolinkurl{arXiv:2012.10367}}].

\bibitem{Brown:1992br}
J.~D. Brown and J.~W. York, Jr., \emph{{Quasilocal energy and conserved charges
  derived from the gravitational action}},
  \href{http://dx.doi.org/10.1103/PhysRevD.47.1407}{\emph{Phys. Rev. D} {\bf
  47} (1993) 1407--1419}.
\newblock
  [\href{https://arxiv.org/abs/gr-qc/9209012}{\nolinkurl{arXiv:gr-qc/9209012}}].

\bibitem{Wald:1999wa}
R.~M. Wald and A.~Zoupas, \emph{{A General definition of `conserved quantities'
  in general relativity and other theories of gravity}},
  \href{http://dx.doi.org/10.1103/PhysRevD.61.084027}{\emph{Phys. Rev. D} {\bf
  61} (2000) 084027}.
\newblock
  [\href{https://arxiv.org/abs/gr-qc/9911095}{\nolinkurl{arXiv:gr-qc/9911095}}].

\bibitem{Chandrasekaran2021}
V.~Chandrasekaran, {\'E}.~{\'E}. Flanagan, I.~Shehzad and A.~J. Speranza,
  \emph{A general framework for gravitational charges and holographic
  renormalization},
  [\href{https://arxiv.org/abs/2111.11974}{\nolinkurl{arXiv:2111.11974}}].

\bibitem{Balasubramanian:1999re}
V.~Balasubramanian and P.~Kraus, \emph{{A Stress tensor for Anti-de Sitter
  gravity}}, \href{http://dx.doi.org/10.1007/s002200050764}{\emph{Commun. Math.
  Phys.} {\bf 208} (1999) 413--428}.
\newblock
  [\href{https://arxiv.org/abs/hep-th/9902121}{\nolinkurl{arXiv:hep-th/9902121}}].

\bibitem{Myers:1999psa}
R.~C. Myers, \emph{{Stress tensors and Casimir energies in the AdS / CFT
  correspondence}},
  \href{http://dx.doi.org/10.1103/PhysRevD.60.046002}{\emph{Phys. Rev. D} {\bf
  60} (1999) 046002}.
\newblock
  [\href{https://arxiv.org/abs/hep-th/9903203}{\nolinkurl{arXiv:hep-th/9903203}}].

\bibitem{DeHaro2001}
S.~{De Haro}, K.~Skenderis and S.~N. Solodukhin, \emph{{Holographic
  reconstruction of spacetime and renormalization in the AdS/CFT
  correspondence}},
  \href{http://dx.doi.org/10.1007/s002200100381}{\emph{Communications in
  Mathematical Physics} {\bf 217} (2001) 595--622}.
\newblock
  [\href{https://arxiv.org/abs/hep-th/0002230}{\nolinkurl{arXiv:hep-th/0002230}}].

\bibitem{Henneaux1979}
M.~Henneaux, \emph{{Zero Hamiltonian signature spacetimes}}, {\emph{Bull. Soc.
  Math. Belg. Ser. A} {\bf 31} (1979) 47--63}.

\bibitem{Duval:2014lpa}
C.~Duval, G.~W. Gibbons and P.~A. Horvathy, \emph{{Conformal Carroll groups}},
  \href{http://dx.doi.org/10.1088/1751-8113/47/33/335204}{\emph{J. Phys. A}
  {\bf 47} (2014) 335204}.
\newblock
  [\href{https://arxiv.org/abs/1403.4213}{\nolinkurl{arXiv:1403.4213}}].

\bibitem{Duval:2014uoa}
C.~Duval, G.~W. Gibbons, P.~A. Horvathy and P.~M. Zhang, \emph{{Carroll versus
  Newton and Galilei: two dual non-Einsteinian concepts of time}},
  \href{http://dx.doi.org/10.1088/0264-9381/31/8/085016}{\emph{Class. Quant.
  Grav.} {\bf 31} (2014) 085016}.
\newblock
  [\href{https://arxiv.org/abs/1402.0657}{\nolinkurl{arXiv:1402.0657}}].

\bibitem{Duval:2014uva}
C.~Duval, G.~W. Gibbons and P.~A. Horvathy, \emph{{Conformal Carroll groups and
  BMS symmetry}},
  \href{http://dx.doi.org/10.1088/0264-9381/31/9/092001}{\emph{Class. Quant.
  Grav.} {\bf 31} (2014) 092001}.
\newblock
  [\href{https://arxiv.org/abs/1402.5894}{\nolinkurl{arXiv:1402.5894}}].

\bibitem{Henneaux:2021yzg}
M.~Henneaux and P.~Salgado-Rebolledo, \emph{{Carroll contractions of
  Lorentz-invariant theories}},
  \href{http://dx.doi.org/10.1007/JHEP11(2021)180}{\emph{JHEP} {\bf 11} (2021)
  180}.
\newblock
  [\href{https://arxiv.org/abs/2109.06708}{\nolinkurl{arXiv:2109.06708}}].

\bibitem{Parattu_2016}
K.~Parattu, S.~Chakraborty, B.~R. Majhi and T.~Padmanabhan, \emph{A boundary
  term for the gravitational action with null boundaries},
  \href{http://dx.doi.org/10.1007/s10714-016-2093-7}{\emph{Gen. Relativ.
  Gravitation} {\bf 48} (2016) }.
\newblock
  [\href{https://arxiv.org/abs/1501.01053}{\nolinkurl{arXiv:1501.01053}}].

\bibitem{Lehner_2016}
L.~Lehner, R.~C. Myers, E.~Poisson and R.~D. Sorkin, \emph{Gravitational action
  with null boundaries},
  \href{http://dx.doi.org/10.1103/physrevd.94.084046}{\emph{Phys. Rev. D} {\bf
  94} (2016) }.
\newblock
  [\href{https://arxiv.org/abs/1609.00207}{\nolinkurl{arXiv:1609.00207}}].

\bibitem{Hopfmuller2017a}
F.~Hopfm{\"{u}}ller and L.~Freidel, \emph{{Gravity degrees of freedom on a null
  surface}}, \href{http://dx.doi.org/10.1103/PhysRevD.95.104006}{\emph{Phys.
  Rev. D} {\bf 95} (2017) 104006}.
\newblock
  [\href{https://arxiv.org/abs/1611.03096}{\nolinkurl{arXiv:1611.03096}}].

\bibitem{Oliveri:2019gvm}
R.~Oliveri and S.~Speziale, \emph{{Boundary effects in General Relativity with
  tetrad variables}},
  \href{http://dx.doi.org/10.1007/s10714-020-02733-8}{\emph{Gen. Rel. Grav.}
  {\bf 52} (2020) 83}.
\newblock
  [\href{https://arxiv.org/abs/1912.01016}{\nolinkurl{arXiv:1912.01016}}].

\bibitem{Aghapour:2018icu}
S.~Aghapour, G.~Jafari and M.~Golshani, \emph{{On variational principle and
  canonical structure of gravitational theory in double-foliation formalism}},
  \href{http://dx.doi.org/10.1088/1361-6382/aaef9e}{\emph{Class. Quant. Grav.}
  {\bf 36} (2019) 015012}.
\newblock
  [\href{https://arxiv.org/abs/1808.07352}{\nolinkurl{arXiv:1808.07352}}].

\bibitem{Jafari:2019bpw}
G.~Jafari, \emph{{Stress Tensor on Null Boundaries}},
  \href{http://dx.doi.org/10.1103/PhysRevD.99.104035}{\emph{Phys. Rev. D} {\bf
  99} (2019) 104035}.
\newblock
  [\href{https://arxiv.org/abs/1901.04054}{\nolinkurl{arXiv:1901.04054}}].

\bibitem{Donnay2019}
L.~Donnay and C.~Marteau, \emph{{Carrollian Physics at the Black Hole
  Horizon}}, \href{http://dx.doi.org/10.1088/1361-6382/ab2fd5}{\emph{Class.
  Quant. Grav.} {\bf 36} (2019) 165002}.
\newblock
  [\href{https://arxiv.org/abs/1903.09654}{\nolinkurl{arXiv:1903.09654}}].

\bibitem{Mars1993}
M.~Mars and J.~M.~M. Senovilla, \emph{{Geometry of general hypersurfaces in
  space-time: Junction conditions}},
  \href{http://dx.doi.org/10.1088/0264-9381/10/9/026}{\emph{Class. Quant.
  Grav.} {\bf 10} (1993) 1865--1897}.
\newblock
  [\href{https://arxiv.org/abs/gr-qc/0201054}{\nolinkurl{arXiv:gr-qc/0201054}}].

\bibitem{Geroch-asymp}
R.~Geroch, {\emph{{Asymptotic structure of space-time}}, } in \emph{{Asymptotic
  structure of space-time}} (F.~P. Esposito and L.~Witten, eds.).
\newblock Plenum Press, New York, 1977.

\bibitem{Ashtekar:1987tt}
A.~Ashtekar, {\emph{{Asymptotic Quantization: based on 1984 Naples lectures}}}.
\newblock Bibliopolis, 1987.

\bibitem{CFP}
V.~Chandrasekaran, {\'{E}}.~Flanagan and K.~Prabhu, \emph{{Symmetries and
  charges of general relativity at null boundaries}},
  \href{http://dx.doi.org/10.1007/JHEP11(2018)125}{\emph{J. High Energy Phys.}
  {\bf 2018} (2018) 1--68}.
\newblock
  [\href{https://arxiv.org/abs/1807.11499}{\nolinkurl{arXiv:1807.11499}}].

\bibitem{Dautcourt1967}
G.~Daŭtcourt, \emph{{Characteristic Hypersurfaces in General Relativity. I}},
  \href{http://dx.doi.org/10.1063/1.1705385}{\emph{Journal of Mathematical
  Physics} {\bf 8} (1967) 1492--1501}.

\bibitem{Gourgoulhon:2005ng}
E.~Gourgoulhon and J.~L. Jaramillo, \emph{{A 3+1 perspective on null
  hypersurfaces and isolated horizons}},
  \href{http://dx.doi.org/10.1016/j.physrep.2005.10.005}{\emph{Phys. Rept.}
  {\bf 423} (2006) 159--294}.
\newblock
  [\href{https://arxiv.org/abs/gr-qc/0503113}{\nolinkurl{arXiv:gr-qc/0503113}}].

\bibitem{Damour1982}
T.~{Damour}, {\emph{{Surface Effects in Black-Hole Physics}}, } in \emph{Marcel
  Grossmann Meeting: General Relativity}, p.~587, 1982.

\bibitem{Burnett1990}
G.~A. Burnett and R.~M. Wald, \emph{{A conserved current for perturbations of
  Einstein-Maxwell space-times}},
  \href{http://dx.doi.org/10.1098/rspa.1990.0080}{\emph{Proceedings of the
  Royal Society of London. Series A: Mathematical and Physical Sciences} {\bf
  430} (1990) 57--67}.

\bibitem{GPS}
A.~M. Grant, K.~Prabhu and I.~Shehzad, \emph{{The Wald-Zoupas prescription for
  asymptotic charges at null infinity in general relativity}},
  [\href{https://arxiv.org/abs/2105.05919}{\nolinkurl{arXiv:2105.05919}}].

\bibitem{W-closed}
R.~M. {Wald}, \emph{{On identically closed forms locally constructed from a
  field}}, \href{http://dx.doi.org/10.1063/1.528839}{\emph{Journal of
  Mathematical Physics} {\bf 31} (1990) 2378--2384}.

\bibitem{Witten:1998qj}
E.~Witten, \emph{{Anti-de Sitter space and holography}},
  \href{http://dx.doi.org/10.4310/ATMP.1998.v2.n2.a2}{\emph{Adv. Theor. Math.
  Phys.} {\bf 2} (1998) 253--291}.
\newblock
  [\href{https://arxiv.org/abs/hep-th/9802150}{\nolinkurl{arXiv:hep-th/9802150}}].

\bibitem{MM-term}
R.~B. Mann and D.~Marolf, \emph{{Holographic renormalization of asymptotically
  flat spacetimes}},
  \href{http://dx.doi.org/10.1088/0264-9381/23/9/010}{\emph{Class. Quant.
  Grav.} {\bf 23} (2006) 2927--2950}.
\newblock
  [\href{https://arxiv.org/abs/hep-th/0511096}{\nolinkurl{arXiv:hep-th/0511096}}].

\bibitem{KS2011}
I.~Kosmann-Schwarzbach,
  \href{http://dx.doi.org/10.1007/978-0-387-87868-3}{\emph{The Noether
  Theorems, Invariance and Conservation Laws in the Twentieth Century}}.
\newblock Springer-Verlag New York, 2011.

\bibitem{Jacobson:2011cc}
T.~Jacobson, \emph{{Initial value constraints with tensor matter}},
  \href{http://dx.doi.org/10.1088/0264-9381/28/24/245011}{\emph{Class. Quant.
  Grav.} {\bf 28} (2011) 245011}.
\newblock
  [\href{https://arxiv.org/abs/1108.1496}{\nolinkurl{arXiv:1108.1496}}].

\bibitem{Seifert:2006kv}
M.~D. Seifert and R.~M. Wald, \emph{{A General variational principle for
  spherically symmetric perturbations in diffeomorphism covariant theories}},
  \href{http://dx.doi.org/10.1103/PhysRevD.75.084029}{\emph{Phys. Rev. D} {\bf
  75} (2007) 084029}.
\newblock
  [\href{https://arxiv.org/abs/gr-qc/0612121}{\nolinkurl{arXiv:gr-qc/0612121}}].

\bibitem{Henningson1998a}
M.~Henningson and K.~Skenderis, \emph{{The holographic Weyl anomaly}},
  \href{http://dx.doi.org/10.1088/1126-6708/1998/07/023}{\emph{J. High Energy
  Phys.} {\bf 1998} (1998) 23}.
\newblock
  [\href{https://arxiv.org/abs/hep-th/9806087}{\nolinkurl{arXiv:hep-th/9806087}}].

\bibitem{Hollands:2005wt}
S.~Hollands, A.~Ishibashi and D.~Marolf, \emph{{Comparison between various
  notions of conserved charges in asymptotically AdS-spacetimes}},
  \href{http://dx.doi.org/10.1088/0264-9381/22/14/004}{\emph{Class. Quant.
  Grav.} {\bf 22} (2005) 2881--2920}.
\newblock
  [\href{https://arxiv.org/abs/hep-th/0503045}{\nolinkurl{arXiv:hep-th/0503045}}].

\bibitem{deBoer:2017ing}
J.~de~Boer, J.~Hartong, N.~A. Obers, W.~Sybesma and S.~Vandoren, \emph{{Perfect
  Fluids}}, \href{http://dx.doi.org/10.21468/SciPostPhys.5.1.003}{\emph{SciPost
  Phys.} {\bf 5} (2018) 003}.
\newblock
  [\href{https://arxiv.org/abs/1710.04708}{\nolinkurl{arXiv:1710.04708}}].

\bibitem{Bekaert:2015xua}
X.~Bekaert and K.~Morand, \emph{{Connections and dynamical trajectories in
  generalised Newton-Cartan gravity II. An ambient perspective}},
  \href{http://dx.doi.org/10.1063/1.5030328}{\emph{J. Math. Phys.} {\bf 59}
  (2018) 072503}.
\newblock
  [\href{https://arxiv.org/abs/1505.03739}{\nolinkurl{arXiv:1505.03739}}].

\bibitem{Ciambelli2019b}
L.~Ciambelli, R.~G. Leigh, C.~Marteau and P.~M. Petropoulos, \emph{{Carroll
  structures, null geometry, and conformal isometries}},
  \href{http://dx.doi.org/10.1103/PhysRevD.100.046010}{\emph{Phys. Rev. D} {\bf
  100} (2019) 046010}.
\newblock
  [\href{https://arxiv.org/abs/1905.02221}{\nolinkurl{arXiv:1905.02221}}].

\bibitem{Bagchi:2019xfx}
A.~Bagchi, A.~Mehra and P.~Nandi, \emph{{Field Theories with Conformal
  Carrollian Symmetry}},
  \href{http://dx.doi.org/10.1007/JHEP05(2019)108}{\emph{JHEP} {\bf 05} (2019)
  108}.
\newblock
  [\href{https://arxiv.org/abs/1901.10147}{\nolinkurl{arXiv:1901.10147}}].

\bibitem{Hartong:2015usd}
J.~Hartong, \emph{{Holographic Reconstruction of 3D Flat Space-Time}},
  \href{http://dx.doi.org/10.1007/JHEP10(2016)104}{\emph{JHEP} {\bf 10} (2016)
  104}.
\newblock
  [\href{https://arxiv.org/abs/1511.01387}{\nolinkurl{arXiv:1511.01387}}].

\bibitem{Ciambelli:2018ojf}
L.~Ciambelli and C.~Marteau, \emph{{Carrollian conservation laws and Ricci-flat
  gravity}}, \href{http://dx.doi.org/10.1088/1361-6382/ab0d37}{\emph{Class.
  Quant. Grav.} {\bf 36} (2019) 085004}.
\newblock
  [\href{https://arxiv.org/abs/1810.11037}{\nolinkurl{arXiv:1810.11037}}].

\bibitem{Ciambelli:2018xat}
L.~Ciambelli, C.~Marteau, A.~C. Petkou, P.~M. Petropoulos and K.~Siampos,
  \emph{{Covariant Galilean versus Carrollian hydrodynamics from relativistic
  fluids}}, \href{http://dx.doi.org/10.1088/1361-6382/aacf1a}{\emph{Class.
  Quant. Grav.} {\bf 35} (2018) 165001}.
\newblock
  [\href{https://arxiv.org/abs/1802.05286}{\nolinkurl{arXiv:1802.05286}}].

\bibitem{Ciambelli:2018wre}
L.~Ciambelli, C.~Marteau, A.~C. Petkou, P.~M. Petropoulos and K.~Siampos,
  \emph{{Flat holography and Carrollian fluids}},
  \href{http://dx.doi.org/10.1007/JHEP07(2018)165}{\emph{JHEP} {\bf 07} (2018)
  165}.
\newblock
  [\href{https://arxiv.org/abs/1802.06809}{\nolinkurl{arXiv:1802.06809}}].

\bibitem{Brown:1996bw}
J.~D. Brown, S.~R. Lau and J.~W. York, Jr., \emph{{Energy of isolated systems
  at retarded times as the null limit of quasilocal energy}},
  \href{http://dx.doi.org/10.1103/PhysRevD.55.1977}{\emph{Phys. Rev. D} {\bf
  55} (1997) 1977--1984}.
\newblock
  [\href{https://arxiv.org/abs/gr-qc/9609057}{\nolinkurl{arXiv:gr-qc/9609057}}].

\bibitem{Booth:2001gx}
I.~S. Booth, \emph{{Metric based Hamiltonians, null boundaries, and isolated
  horizons}}, \href{http://dx.doi.org/10.1088/0264-9381/18/20/305}{\emph{Class.
  Quant. Grav.} {\bf 18} (2001) 4239--4264}.
\newblock
  [\href{https://arxiv.org/abs/gr-qc/0105009}{\nolinkurl{arXiv:gr-qc/0105009}}].

\bibitem{Hopfmuller2018}
F.~Hopfm{\"{u}}ller and L.~Freidel, \emph{{Null conservation laws for
  gravity}}, \href{http://dx.doi.org/10.1103/PhysRevD.97.124029}{\emph{Phys.
  Rev. D} {\bf 97} (2018) 124029}.
\newblock
  [\href{https://arxiv.org/abs/1802.06135}{\nolinkurl{arXiv:1802.06135}}].

\bibitem{Wald:1993nt}
R.~M. Wald, \emph{{Black hole entropy is the Noether charge}},
  \href{http://dx.doi.org/10.1103/PhysRevD.48.R3427}{\emph{Phys. Rev. D} {\bf
  48} (1993) R3427--R3431}.
\newblock
  [\href{https://arxiv.org/abs/gr-qc/9307038}{\nolinkurl{arXiv:gr-qc/9307038}}].

\bibitem{Bhattacharyya:2007vjd}
S.~Bhattacharyya, V.~E. Hubeny, S.~Minwalla and M.~Rangamani, \emph{{Nonlinear
  Fluid Dynamics from Gravity}},
  \href{http://dx.doi.org/10.1088/1126-6708/2008/02/045}{\emph{JHEP} {\bf 02}
  (2008) 045}.
\newblock
  [\href{https://arxiv.org/abs/0712.2456}{\nolinkurl{arXiv:0712.2456}}].

\bibitem{Rangamani:2009xk}
M.~Rangamani, \emph{{Gravity and Hydrodynamics: Lectures on the fluid-gravity
  correspondence}},
  \href{http://dx.doi.org/10.1088/0264-9381/26/22/224003}{\emph{Class. Quant.
  Grav.} {\bf 26} (2009) 224003}.
\newblock
  [\href{https://arxiv.org/abs/0905.4352}{\nolinkurl{arXiv:0905.4352}}].

\bibitem{Penna:2017vms}
R.~F. Penna, \emph{{BMS$_3$ invariant fluid dynamics at null infinity}},
  \href{http://dx.doi.org/10.1088/1361-6382/aaa3aa}{\emph{Class. Quant. Grav.}
  {\bf 35} (2018) 044002}.
\newblock
  [\href{https://arxiv.org/abs/1708.08470}{\nolinkurl{arXiv:1708.08470}}].

\bibitem{Ciambelli:2020eba}
L.~Ciambelli, C.~Marteau, P.~M. Petropoulos and R.~Ruzziconi, \emph{{Gauges in
  Three-Dimensional Gravity and Holographic Fluids}},
  \href{http://dx.doi.org/10.1007/JHEP11(2020)092}{\emph{JHEP} {\bf 11} (2020)
  092}.
\newblock
  [\href{https://arxiv.org/abs/2006.10082}{\nolinkurl{arXiv:2006.10082}}].

\bibitem{Fareghbal:2013ifa}
R.~Fareghbal and A.~Naseh, \emph{{Flat-Space Energy-Momentum Tensor from
  BMS/GCA Correspondence}},
  \href{http://dx.doi.org/10.1007/JHEP03(2014)005}{\emph{JHEP} {\bf 03} (2014)
  005}.
\newblock
  [\href{https://arxiv.org/abs/1312.2109}{\nolinkurl{arXiv:1312.2109}}].

\bibitem{Kapec:2016jld}
D.~Kapec, P.~Mitra, A.-M. Raclariu and A.~Strominger, \emph{{2D Stress Tensor
  for 4D Gravity}},
  \href{http://dx.doi.org/10.1103/PhysRevLett.119.121601}{\emph{Phys. Rev.
  Lett.} {\bf 119} (2017) 121601}.
\newblock
  [\href{https://arxiv.org/abs/1609.00282}{\nolinkurl{arXiv:1609.00282}}].

\bibitem{FN}
{\'E}.~{\'E}. Flanagan and D.~A. Nichols, \emph{{Conserved charges of the
  extended Bondi-Metzner-Sachs algebra}},
  \href{http://dx.doi.org/10.1103/PhysRevD.95.044002}{\emph{Phys. Rev.} {\bf
  D95} (2017) 044002}.
\newblock
  [\href{https://arxiv.org/abs/1510.03386}{\nolinkurl{arXiv:1510.03386}}].

\bibitem{Compere:2019gft}
G.~Comp\`ere, R.~Oliveri and A.~Seraj, \emph{{The Poincar\'e and BMS
  flux-balance laws with application to binary systems}},
  \href{http://dx.doi.org/10.1007/JHEP10(2020)116}{\emph{JHEP} {\bf 10} (2020)
  116}.
\newblock
  [\href{https://arxiv.org/abs/1912.03164}{\nolinkurl{arXiv:1912.03164}}].

\bibitem{Chakraborty2019}
S.~{Chakraborty} and K.~{Parattu}, \emph{{Null boundary terms for
  Lanczos-Lovelock gravity}},
  \href{http://dx.doi.org/10.1007/s10714-019-2502-9}{\emph{General Relativity
  and Gravitation} {\bf 51} (2019) 23}.
\newblock
  [\href{https://arxiv.org/abs/1806.08823}{\nolinkurl{arXiv:1806.08823}}].

\bibitem{Vogel1965}
W.~Vogel, \emph{{\"Uber lineare Zusammenh\"ange in singul\"aren Riemannschen
  R\"aumen}}, \href{http://dx.doi.org/10.1007/BF01220008}{\emph{Arch. Math.}
  {\bf 16} (1965) 106--116}.

\bibitem{Leigh-notes}
L.~Ciambelli and R.~G. Leigh. Unpublished notes.

\end{thebibliography}\endgroup
\end{document}